# PAH chemistry at eV internal energies.

# 2. Ring alteration and dissociation

Georges Trinquier, Aude Simon, Mathias Rapacioli, Florent Xavier Gadéa

Laboratoire de Chimie et Physique Quantiques (CNRS, UMR5626),
IRSAMC, Université Paul-Sabatier (Toulouse III), 31062 Toulouse Cedex, France

**Abstract**. Recognized as important interstellar constituents, polycyclic aromatic hydrocarbons (PAHs) have been intensively studied in astrochemistry and their spectroscopy, thermodynamics, dynamics, and fragmentations are now amply documented. There exists typical alternatives to the ground-state regular planar structures of PAHs, as long as they bear internal energies in the range 1-10 eV. Resulting from intramolecular rearrangements, such high-lying minima on the potential-energy surfaces should be taken into consideration in the studies of PAH processing in astrophysical conditions. Resting upon DFT calculations mainly performed on two emblematic PAH representatives, coronene and pyrene, in their neutral and mono- and multi-cationic states, this second survey addresses the following alternatives: (1) opened forms containing ethynyl or 2-butynyl groups, (2) vinylidene isomers, in which phenanthrene patterns are reorganized into dibenzofulvene ones, (3) "twisted" forms, where external CH=CH bonds can be partly twisted, and (4) bicyclobutane forms, in which the latter are integrated in saturated bicyclic forms. A few scenarios for elimination of fragments H, $H_2$, $C_2H_2$ and $C_2H_4$ are explored. As far as possible, familiar concepts of organic chemistry, such as aromaticity or Clar's rules, are invoked for interpretations.

# 1. Introduction

Since the presence of polycyclic aromatic hydrocarbons (PAHs) in the interstellar medium has been proposed in the mid-eighties (Allamandola, Tielens, & Barker 1985; Léger & Puget 1984), their interest has been constantly reinforced for years. As astro-PAHs are supposed to be the carriers of the aromatic infrared bands (AIBs), extensive experimental and theoretical efforts have been devoted in trying to assign specific carriers, although no molecule has yet been unambiguously identified (Joblin & Tielens 2011). Arising from natural processes, PAHs are also present in the atmosphere, as efficiently formed byproducts (Finlayson-Pitts & Pitts 1987). On another hand, they are also well-known in organic chemistry (Berionni, Wu, & Schleyer 2014; Bodwell 2014; Ciesielski et al. 2006; Hopf 2014; Muller, Kubel, & Mullen 1998; Popov & Boldyrev 2012; Scott 1982), where ground-state low-energy configurations and conformations - typically within a few kcal/mol or 0.05 eV - have been addressed thoroughly. Yet in many astrophysical and terrestrial situations, PAH systems are hot, carrying internal energies of several eV. The goal of the present study is to explore theoretically PAH chemistry at these rather high internal energies, more familiar to physicists than to chemists, while trying to keep advantage of the current chemical knowledge - a set of simple concepts, laws and rules that characterize "chemical intuition".

In the first part of this work (Trinquier et al. 2017), we focused on hydrogen shifting along the carbon skeleton, leading to various H-shifted isomers. This systematic exploration sometimes pointed to ring opening into extracyclic ethynyl derivatives. In the present contribution, we now explore a large set of structural isomers built from ring alterations of various types, with further considerations on scrambling and dissociation here referring to the complete set of explored structures. As in the preceding companion paper, we mainly focus on emblematic PAHs coronene and pyrene in their neutral, and mono-, di-, and tri-cationic states and we look for general trends of PAHs departing from their normal regular form and remaining in their initial multiplicity state.

The paper is organized as follows. Section 2 briefly recalls the computational details. In Section 3, we will address the topological alteration of selected ring opening with concomitant creation of C≡C triple bonds. Section 4 is devoted to another local topological alteration, namely the formation of frames including five-membered rings bearing extracyclic vinylidene groups (dibenzofulvene forms). Next, more disrupting alterations will be addressed, such as local twists at one or several -HC-CH- units (Section 5), and bicyclobutane-carrying forms (Section 6). Last, in Section 7 we will recapitulate how these various structural isomers are interconnected, how they can give hydrogen or carbon exchange, and we will discuss some fragmentation pathways.

**2. Methods and computational details.**

As in the previous companion paper, we focus here on the potential energy surfaces (PES) associated to the ground state of PAHs in their normal forms. Therefore, this study will essentially limit to closed-shell *singlet* surfaces for neutral and dicationic species, and to *doublet* surfaces for monocationic and tricationic species, as these multiplicities correspond in general to the ground state of neutral and charged species in their normal regular forms. Occasionally, triplet or diradical singlet states are also examined, but their exhaustive and systematic study is beyond the scope of the present work, and should be the object of a separate work.

All species are treated within density-functional-theory approach, in restricted (DFT) or unrestricted (UDFT) modes, using standard B3LYP hybrid functional with 6-311G** basis sets, as implemented in the *Gaussian09* quantum chemistry package (Frisch et al.). Full geometry optimizations were carried out, free of any symmetry constraint, up to energy gradients lower than $10^{-5}$ au. The natures of the so-obtained stationary points (minima or saddle-point) were determined through Hessian diagonalizations and vibrational analyses is systematically performed on the so-obtained stationary points in order to characterize their nature. When given, frequencies correspond to uncorrected values. Likewise, no scaling factor is applied in the evaluation of zero-point vibrational energy (ZPE) corrections.

For structures carrying one or several C-CH-CH-C twisted sequences addressed in Section 5, we trust more complete-active-space-SCF (CASSCF) descriptions rather than DFT ones, as these systems are multiconfigurational by essence. These species were therefore the object of small CASSCF treatments, namely CAS(2,2), CAS(4,4) and CAS(6,6), in parallel with the DFT treatment. For a succinct description of Clar's rules, the reader is referred to Section 2 of the previous companion paper.

**3. Ring opening into allenic or triply-bonded structures.**

As mentioned in the previous companion paper, a strong attractor associated to the existence of extracyclic ethynyl group –C≡CH has been seen to interfere in the travel of hydrogen along the rim. Such a structure happens to be fairly low in energy, at only 2.71 eV (2.60 eV after ZPE correction) above the normal forms of coronene and at 2.47 eV (ZPE, 2.36 eV) above the normal form of pyrene. We examine such alternatives in this section. There are three main possibilities of ring opening in PAHs.

*A- Allenic derivatives*. Simple breaking of a given radialenic-type bond *without any hydrogen migration* leads to di-allenic ten-membered ring structures (Scheme 1, top). Because of the orthogonal layout of allenic ends, such an arrangement is very unlikely in a local context of phenalenic frame (Scheme 1, bottom). This opening option is therefore not offered to most polycondensed 'true' PAHs, while being accessible to polyacenes, phenanthrene, triphenylene, tetrabenzo-naphtalene, etc., as illustrated in Appendix A. We have explored some of these cases, namely, naphthalene, phenanthrene and tetrabenzo-naphthalene. For naphthalene, the di-allenic ten-membered ring isomer lies at 4.5 eV (ZPE, 4.4 eV) above the normal form, with a barrier of 5.5 eV (ZPE, 5.2 eV) to overcome for breaking the central C-C bond. Appending benzo groups on allowed edges seems to significantly penalize this opened form since its relative energy is raised to 5.3 eV (ZPE, 5.1 eV) in phenanthrene and to 7.0 eV (ZPE, 6.8 eV) in tetrabenzo-naphthalene (see Appendix B).

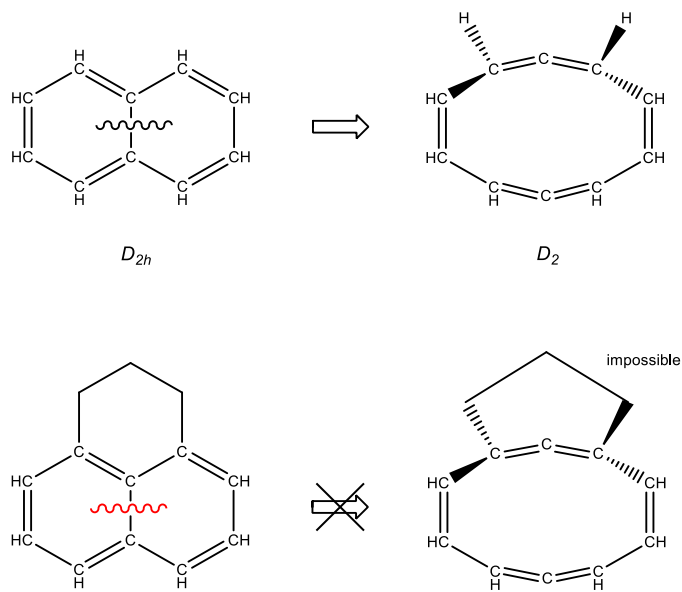

Scheme 1

When hydrogen transfers are allowed, other ways of ring breaking come forward. The two main categories of them associated with single hydrogen relocation are exemplified for our two systems in Scheme 2. When accompanied by appropriate hydrogen migration – here formally of 1→3 type - breaking an external carbon-carbon bond leads to an extracyclic ethynyl derivative, while breaking an internal (or radialenic in coronene) bond leads to a 2-butynyl ten-membered ring derivative. In the latter case, the aligned C–C≡C–C set can be written alternatively under a butatriene cumulenic form C=C=C=C. Which mesomeric form will be favored depends on the position of exterior aromatic rings along the ten-membered ring. Note that these unimolecular rearrangements are possible in simple phenanthrene, as well, although in this case the ethynyl form is a derivative of diphenyl, thus experiencing some free rotation around the central bond. In the other PAHs, as the ethynyl derivatives undergo no major geometrical constraint, Clar's disjoint sextets (Clar 1964, 1972) are preserved, integrally in pyrene or phenanthrene, and only one subset of them in coronene. By contrast, in the 2-butyne derivatives, the non-planar ten-membered ring perturbs the conjugation frame in such a way that there remain only two Clar's sextets in coronene and one in pyrene. Presumably, this will take some part in the relative stabilities of these opened forms. The relative energies for a selection of such structural isomers - using Clar's rules as guideline - are listed in Table 1.

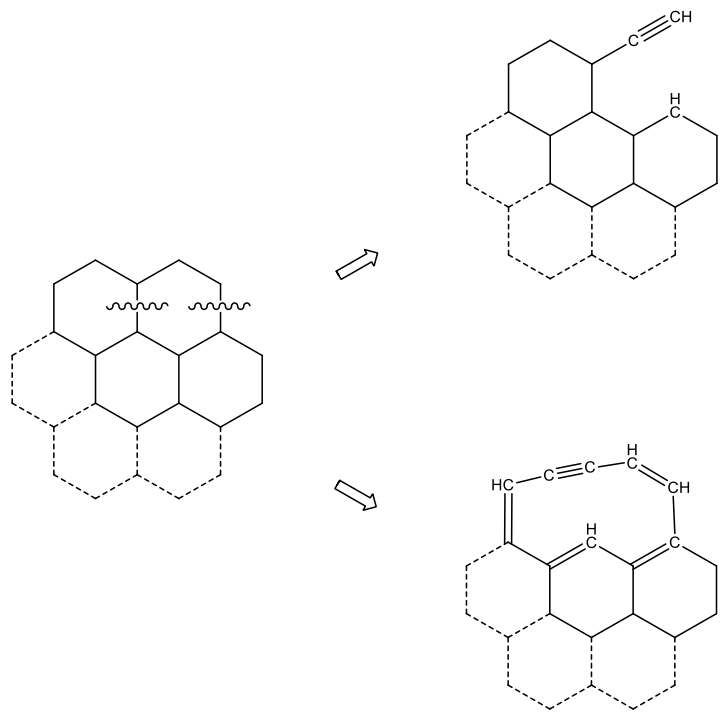

Scheme 2

*B- Extracyclic ethynyl derivatives.* As expected, and as mentioned above, the mono-ethynyl forms lie rather low in energy above the normal forms (2.7 eV and 2.5 eV for coronene and pyrene respectively; ZPE, 2.6 eV and 2.4 eV respectively). For coronene, this form can be reached from 1,2- and 1,3-hydrogen-shifted isomers by means of a barrier of 4.7 eV above the normal form (ZPE, 4.5 eV), as illustrated in Figure 2 of the companion paper. For pyrene, this form can be reached from 1,2- and 1,4- hydrogen-shifted isomers by means of a barrier of 5.4 eV above the normal form (ZPE, 5.2 eV), as illustrated in Figure 4 of the companion paper. Cationic mono-ethynyl forms were also explored and their relative energies remain close to the neutral cases (Table 1). They range at 2.2-2.5 eV for coronene, while at 2.5-3.1 eV for pyrene. The effect of ionization on the relative stability of ethynyl isomers is therefore different in both compounds, and one can note the high value for the dication of pyrene (3.1 eV).

All mono-ethynyl coronenes are found to be non-planar, a slight warping of the skeleton relieving the interaction between the CH and C≡CH groups. However, once ZPE is taken into account, planar forms are either preferred or of equal energy than these distorted forms. In pyrene, all mono-ethynyl forms are found to be planar, except the trication, but again this form is degenerate with the planar form once ZPE is taken into account. In summary, mono-ethynyl structural isomers are all virtually planar with some loosing allowed to skeleton buckling.

As shown in Figure 1, it is possible to open more rings to create several extracyclic ethynyl groups while keeping Clar's aromatic sextets (one of the subset, for coronene). We have explored these structural isomers, and their relative energies are reported in Table 1. Presumably due to this preservation of aromatic sextets, the cost for creating extracyclic ethynyl by ring breaking seems to be a fairly transferable constant increment. This point is further evidenced by making explicit the energy differences between variously-opened derivatives, as detailed in Appendix C. In coronene, the increment is 2.6-2.7 eV (ZPE, 2.5-2.6 eV), while it is 2.1-2.5 eV (ZPE, 2.0-2.4 eV) in pyrene. In the latter case, it can be pointed out that for di-ethynyl forms, we no longer have here a rigid-prone condensed polycyclic system, but only a diphenyl-type one, allowing some relieve of geometrical constraints. This should explain the lower values for the increments in these structures, and to some extent in the ethynyl form of phenanthrene as well. We will see in a forthcoming section that similar transferability also holds in the context of fulvene (or vinylidene) ring alteration, which again does not destroy the layout of Clar's sextets, and which is associated to an increment of about 2 eV.

*C- Intracyclic 2-butyne derivatives*. As anticipated from Scheme 2, ring breaking into 2-butyne derivatives should disrupt the planar layout and the conjugation scheme of the PAH. As such, the 2-butyne derivatives are expected to be less stable than the ethynyl derivatives. Selected cases of them, drawn in Figure 2, have been explored, and their energies are given in the lower part of Table 1. Let us focus, first, on opened structures containing a single 2-butyne sequence. There is one such structure for coronene, while there exist two isomers for pyrene. In all cases, these forms are lying at about 1.3 eV above the ethynyl derivatives. Just as natural pathways to extracyclic ethynyl forms proceed *via* 1,3-type hydrogen-shifted isomers, a natural pathway to 2-butyne forms can consists in simple bond breaking from inner hydrogen-shifted isomers (Scheme 3). Actually, a barrier of only 0.4 eV is found to link the 1,19-hydrogen shifted isomer of coronene to its 2-butyne form, while a barrier as low as 0.1 eV is found to link 4,16-hydrogen shifted isomer of pyrene to its 2-butyne most stable form *l* (see Figure S1 of the supplementary material). The alternative form *m*, lying less than 0.2 eV above in energy, is accessible from the 1,15-hydrogen shifted isomer.

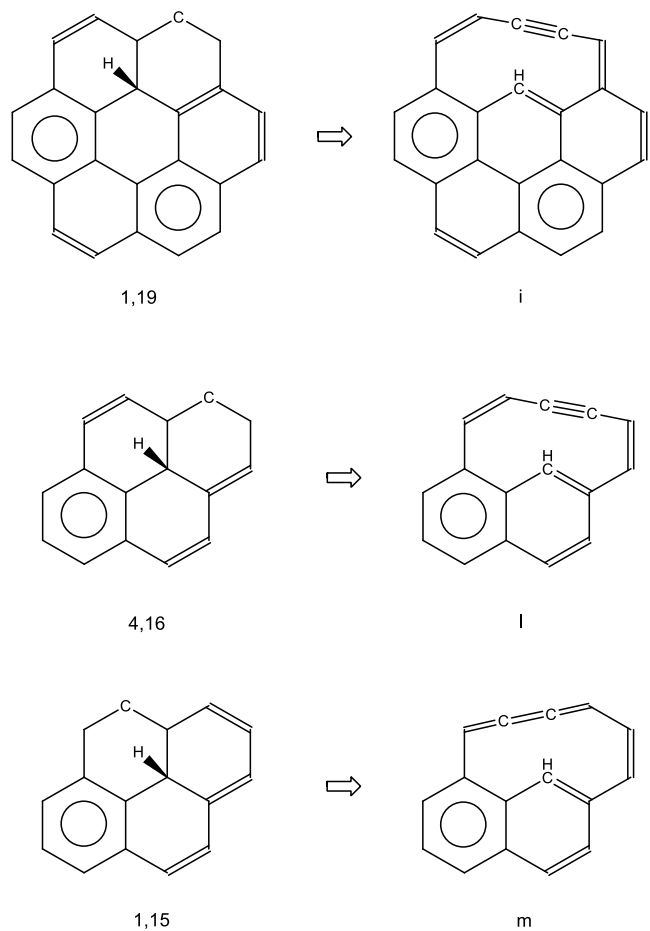

Scheme 3

These butynyl derivatives of pyrene exemplify the way intra-ring sequences –HC–C–C–CH– are best transcribed. Whether it is more appropriate to write a triple bond C–C≡C–C or a cumulenic form C=C=C=C is simply governed by conjugation distribution, given that the two bonds leaving the aromatic ring are necessarily single bonds. Writing a triple bond in isomer *m* would convert an aromatic sextet into a less favored *ortho*-quinonic form. This can be further illustrated in the butynyl derivatives of less symmetric $C_{2h}$ phenanthrene. Formally, there are four possible positions of –C–C– set along the ten-membered-ring, and alternation of both descriptions comes up conspicuously (Scheme 4). The bottom structure, here, is not accessible from simple 1→3 hydrogen migration.

Moreover, despite its higher $C_s$ symmetry, this isomer is not favored geometrically as the ten-membered ring is compelled to substantial puckering, entailing less thermodynamic stability with respect to the three other isomers.

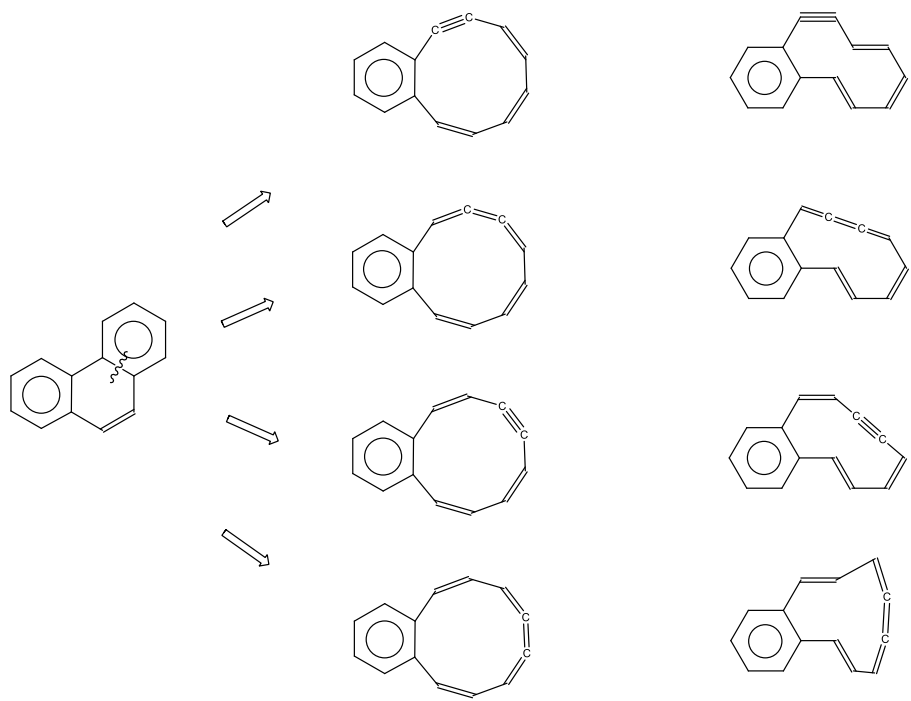

Scheme 4

As can be seen in Table 1, the mono-butyne forms are lying for PAH at about the same height in energy above the normal forms: 4.1 eV for coronene, 3.8-4.0 eV for pyrene, and 3.7-3.8 eV for phenanthrene. As mentioned above, these gaps are significantly larger than those associated to the mono-ethynyl forms (2.7 eV in coronene, 2.5 eV in pyrene, and 2.3 eV in phenanthrene) - clearly, ring opening by breaking inner bonds to produce intra-ring butyne derivatives is a less favored process than exterior opening into extracyclic ethynyl derivatives.

Just as we examined above the transferability of ethynyl energy increments, one can check whether such transferability operates when creating a second or a third butyne function through more ring breaking. Not unexpectedly, the number of possible isomers here expands. In coronene where a single mono-butyne form exists, one can build five di-butyne isomers, and five tri-butyne ones, all drawn in Figure S2 of the supplementary material. In pyrene, two mono-butyne forms exist, and each can generate three di-butyne forms, all drawn in Figure S3 of the supplementary material. This adds up to six di-butynyl isomers for pyrene, more than the five ones existing for coronene. Again, this illustrates how smaller size can be over-compensated by lower symmetry. A few of these species, sketched in Figure 2, have been explored and their energies are reported in the lower part of Table 1. The increment for a second ring opening is fairly conserved in coronene since di-butynyl systems *j* and *k* are at 8.2 eV above the normal form. For pyrene, things are different: the second ring breaking is here less energy demanding since the di-butynyl systems *n* and *o* are lying at 6.8 eV and 5.4 eV respectively, above the normal form. Note the particular stability of isomer *o*, resulting from the breaking of vicinal inner bonds, thus preserving an aromatic sextet.

The geometries of molecules including one or several ten-membered rings are of course non planar, but this occurs at different extent. The single-butyne structures are not far from planarity, while those structures bearing two butyne groups manage to have the remaining aromatic ring at about 45° of the mean plane of the ten-membered rings, as illustrated in Figure 3. In isomer *o*, the $C_{2v}$ transition state corresponding to full orthogonality of the phenyl group with respect to the main large ring lies at less than 0.2 eV (4 kcal/mol) above the equivalent minima. The nature butyne/butatriene of –HC–C–C–CH– sequences can be grasped from optimized CC bond lengths. In our examples, the three-bond length sets all locate in between those of 2-butyne and butatriene limiting forms, with closeness along the lines of conjugation schemes preserving aromatic sextets (see Figure S4 of the supplementary material).

In summary, opening one or several rings in a given PAH gives rise to a lot of possibilities, depending on which cycle is opened and which bond is broken. From the exploration of our modest sampling, ring opening by breaking an outer CC bond into an extracyclic ethynyl derivative is the best favored process, with global exothermicity around 2.5-2-7 eV with respect to normal forms, (ZPE, 2.4-2.6 eV) and overall barriers of 4.7 eV for coronene, and 5.4 eV for pyrene (ZPE, 4.5 eV and 5.2 eV resp.). Opening into a ten-membered ring bearing a 2-butyne motif is less favored thermodynamically (4.1 eV for coronene, 3.8 eV for pyrene), but the overall barrier to be overcome from the normal form is only of 4.8 eV in coronene, and 4.7 eV in pyrene (ZPE, 4.7 eV and 4.6 eV, resp.), which would suggest that in pyrene, this opening is kinetically favored over the former one.

**4. Vinylidene alternatives**.

Seeing our PAHs as sets of Clar's sextets may suggest that replacing an 'interstitial' six-membered ring by a fulvene moiety (or a vinylidene bridge) should not penalize too much its delocalization energy, nor its total energy, as long as geometrical constraints can be distributed over a whole skeleton (Scheme 5). Moreover, putting aside the non-planarity that is required when several of these bridges are installed along a PAH edge, the conservation of Clar's pattern suggests that penalty increments per vinylidene bridge should be conserved upon alterning such substitutions along the rim.

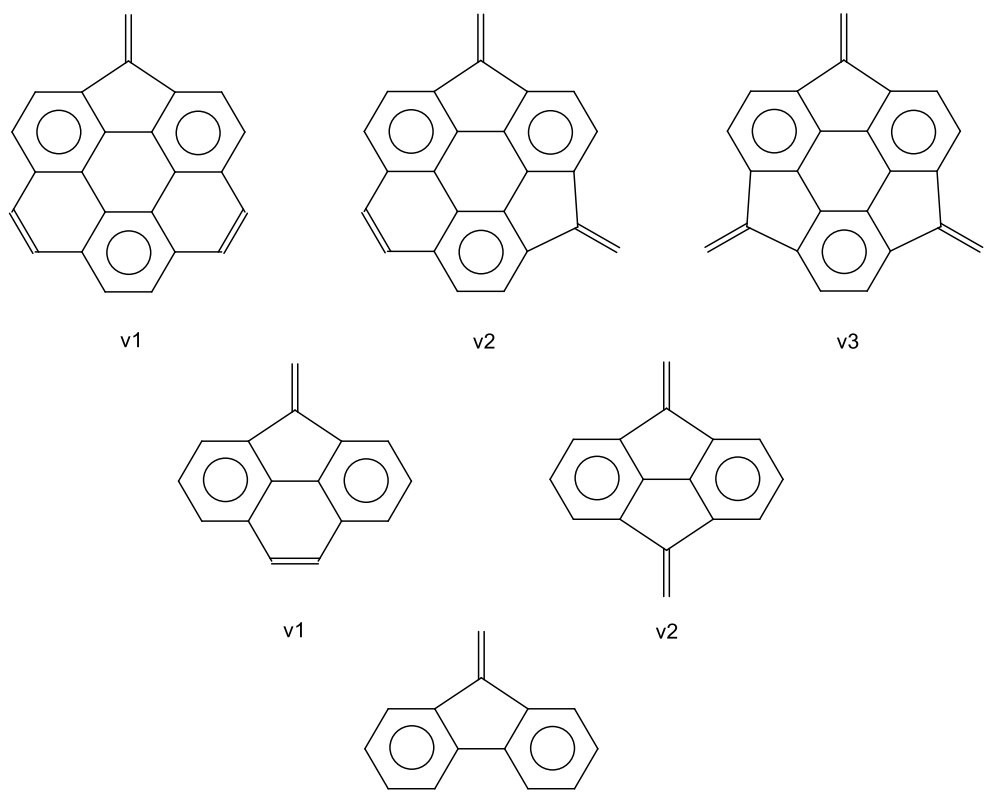

Scheme 5

*A- Thermodynamic stabilities*. The relative energies of various vinylidene alternatives for coronene, pyrene and phenanthrene, are given in Table 2, for neutral and charged species. Again, the case of phenanthrene is special in that here, by construction, planar angular constraints can be released more easily than in larger PAH. This should account why the vinylidene alternative dibenzo-fulvene is lying only 0.69 eV above normal phenanthrene (ZPE, 0.65 eV). The mono-vinylidene alternative of pyrene is lying at 1.16 eV above its normal form, (ZPE, 1.11 eV), while the mono-vinylidene alternative for coronene is lying at 1.60 eV above its normal form (ZPE, 1.55 eV). The three values reflect deformation costs in the right ordering, with a low value for phenanthrene and a high value for coronene. All these *v1* mono-vinylidene derivatives are found to be planar at our level of description.

Increasing the number of vinylidene bridges while maintaining the Clar pattern leads to di-vinylidene stuctures, *v2*, for pyrene and coronene, and a tri-vinylidene structure, *v3*, for coronene (Scheme 5). For both the neutral and cationic species, *v2* is found to be planar for pyrene, and non-planar for coronene, which here adopts a cup-shaped form, more regular and completed in the tri-vinylidene form *v3*, and even more so in the hexa-vinylidene isomer *v6* of circum-coronene discussed hereunder (Figure 4). For coronene, one understands that geometrical constraints induced by a sole five-membered ring can be assimilated throughout the remaining *planar* frame, while incorporating two or more such five-membered rings should inescapably incur curvature. Actually, planar forms of *v2* and *v3* are lying at 2.8 and 21.4 kcal/mol (0.12 and 0.93 eV), respectively, above their relaxed forms. In pyrene *v2*, the pair of five-membered rings is arranged in a different way than it is in coronene *v2*, which should account for its retained planarity - pentalene is not fulvalene.

The relative energies of poly-vinylidene derivatives given in Table 2 would confirm the existence of a reasonably constant and transferable energy increment per vinylidene bridge. For neutral species, the di-vinylidene forms are lying at 3.34 eV and 3.58 eV above the normal forms of coronene and pyrene respectively (ZPE, 3.23 eV and 3.48 eV resp.). The tri-vinylidene form of coronene *v3* is lying at 5.40 eV (ZPE 5.26 eV). Given that non-planar skeletal warping here interferes with delocalization effects, such an observed increment of 1.6-2.0 eV per vinylidene unit supports the view of additive transferable penalties whenever a vinylidene bridge is inserted on backdrop of *unchanged* Clar configuration. The same increment also holds for charged species.

To further check this seeming universality, we have examined the hexa-vinylidene isomer of a larger PAH, namely circum-coronene (Scheme 6). In its spherical-cap relaxed form, the hexa-vinylidene isomer is lying 9.81 eV above normal planar circum-coronene, reducing to a satisfactory unit price of 1.64 eV per vinylidene bridge. These correlations confirm a sensible linear dependence of energy against extracyclic vinylidene units. HMO treatment shows a similar additivity, with an increment here around 0.22 $|\beta|$ per vinylidene unit. It is remarkable but not inconsistent to observe a near additivity in both cases. DFT explicitly describes real systems, subject to geometrical constraints inducing strong departures from planarity (compelling the above hexa-vinylidene molecule to planarity requires as much as 5.2 eV) and conjugation losses. Hückel treatment measures delocalization effects from topological information only, which is also what analyses in term of Clar sextets are all about. Illustrations of both linear dependence are given as supplementary material (Figures S5 and S6). Note that the hexa-vinylidene isomer of circum-coronene has a strikingly-regular spherical shape (Figure 4, bottom), although the constraint arises from the peripherical pentagons. Actually, this system meets Euler's closure condition for making a hemisphere, while a complete closure (fullerene-like) would require incorporation of six additional pentagons.

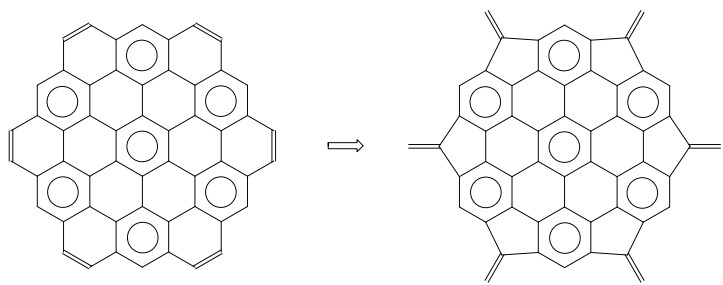

Scheme 6

*B- Routes to formation*. To form a vinylidene isomer from a normal PAH, one can imagine a straight route through direct ring reduction, followed by (or concomitant with) 1→2 hydrogen shift toward extracyclic CH (Scheme 7). Such a route actually exists, even though alternative indirect routes are more energetically favorable, as we shall see below.

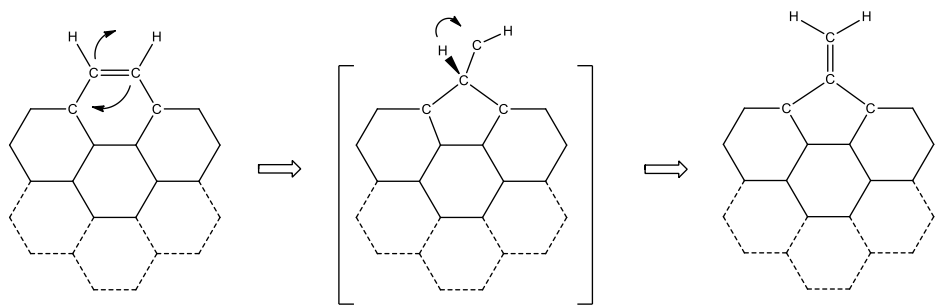

Scheme 7

For neutral states, the barriers associated with this route are lying at 5.3 eV and 4.9 eV above the normal forms of coronene and pyrene, respectively (ZPE, 5.2 and 4.8 eV resp.). For cationic species, these numbers are of the same order (Table 3, top). Actually, for the monocation of coronene, and for the dications of coronene and pyrene, the transitory form in the middle of scheme 7 is a shallow minimum (in coronene monocation, there is even two such minimum conformers). This means that the potential surface around this intermediate structure may be more or less flat. Because such minima could disappear upon changing functional or basis set, we will not further document them, and the barriers selected in Table 3 correspond to the larger ones. For the trication of pyrene, a high-spin solution corresponding to $<S^2>=1.13$ is found at 0.3 kcal/mol only below the normal low-spin doublet state corresponding to $<S^2>=0.78$, warning for possible interference of high spin configurations along the pathways.

An alternative route to vinylidene forms from 1,2-hydrogen shifted isomer is favored energetically, as barriers for the corresponding concerted sigmatropic migrations (Scheme 8) are lower than those discussed above (Table 3, bottom). These transition states are just above those leading to 1,2-hydrogen shifted isomers from normal forms, given in Tables 1 and 3 of the preceding paper (Trinquier *et al.,* 2017), except for coronene dication, pyrene dication and pyrene trication, where they are just below. In any case, the indirect pathway *via* 1,2-hydrogen shifted isomers is always clearly below the direct pathway. Again, for cationic states, the barriers are of the same order than for neutral states. Consistently enough, all barriers for pyrene are lower than those of coronene. Energy profiles for the whole pathways depicted in Scheme 9 for neutral coronene are given in Figure 5. For neutral pyrene, curve shapes are similar, although slightly shifted downwards (see Figure S7 in the supplementary material, which also depicts transition state geometries in Figure S8).

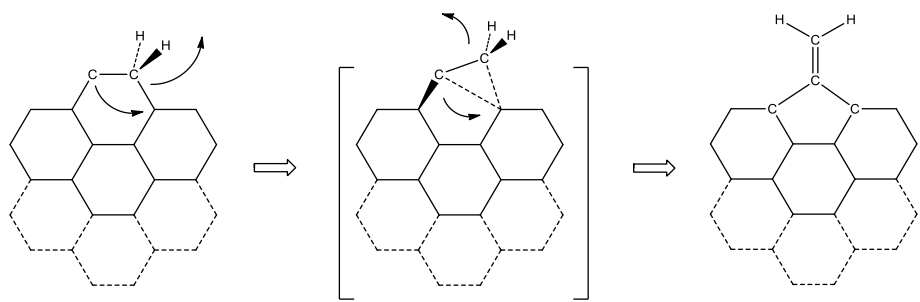

Scheme 8

The trends discussed above also hold for phenanthrene. As the final vinylidene product is here lying at only 0.68 eV above the normal form, the barriers are lower than in the pyrene case. The direct barrier is calculated at 4.86 eV, while the two barriers along the indirect pathway are found at 3.60 eV and 4.00 eV (ZPE, 4.74, 3.46 and 3.91 eV resp.).

The barriers to overcome for reaching poly-vinylidene isomers seem to possess, here also, and to reasonable extent, their own transferability, as can be checked in Appendix D. All barriers for the direct route converge around 5 eV, while those for indirect routes converge around 3.0-3.5 eV for TS1 and 1 eV for TS2 . For neutral direct pathways, this transferability is supported by the imaginary vibrational frequencies, all included in the interval 310-333 cm$^{-1}$. A schematic energy profile for the direct path from neutral normal coronene to its vinylidene forms is given in Figure S9. For neutral pyrene, energy profiles for the two interconversion routes are similar to those of coronene. Given as supplementary material (Scheme S10 and Figure S11), they reveal in this case a poorly-transferable barrier for TS2 - probably due here to ring stress.

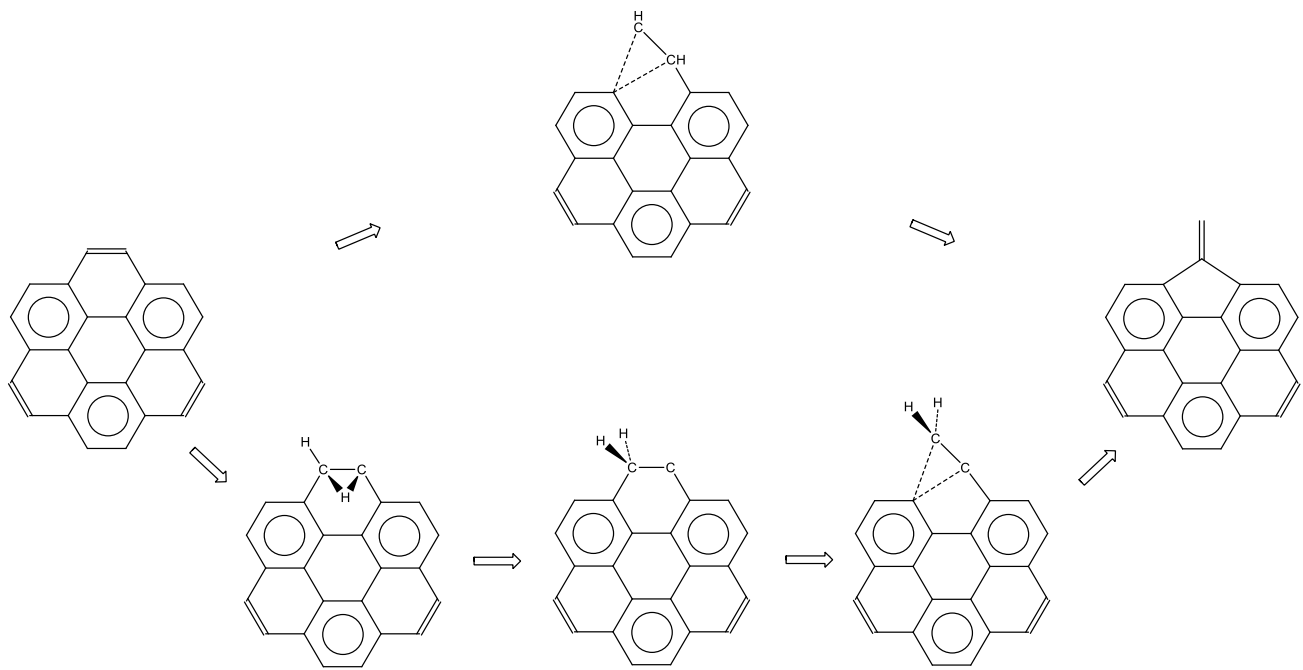

Scheme 9

*C- Comparison with azulene-type alternatives.* As a conclusion to this section, we wish to compare the stability of our vinylidene (or fulvene-type) isomers of PAH to the well-documented azulene-type alternatives made of joined five- and seven-membered rings. For naphthalene, this isomer is known as azulene, while for coronene, the altered isomer with three such sets is known as iso-coronene (we will label them *a* and *a3* respectively) (Ciesielski, et al. 2006; Popov & Boldyrev 2012). The few test cases depicted in Scheme 10 have been examined, and the corresponding results are given in Table 4. At present level of description, benzo-fulvene *bf* is found to be more stable than azulene *a* by 0.4 eV. For a single alteration of coronene, the mono-vinylidene alternative *f1* is still preferred to the mono-azulene one *a1* by 0.2 eV. For two substitutions, *a2* is now more stable than *f2* by 0.3 eV, and iso-coronene *a3* is definitely more stable than the tri-vinylidene isomer *f3* by 1.0 eV. Geometries of *a1*, *a2* and *a3* are all planar, while, as mentioned, those of *f2* and *f3* are bowl-shaped. This suggests that geometrical strains are less severe in iso-coronene *a3* than in the tri-vinylidene form *f3*, making possible for the former to efficiently distribute them all over the skeleton while preserving full planarity - hence maximum delocalization. This no longer holds in the *a6* variant of circum-coronene, iso-circum-coronene, which exhibits the same cup-shaped form as *f6* (see Figure S12 in the supplementary material). Oddly enough, these two forms have virtually the same energy, and incidentally share nearly identical HMO delocalization energies (see Table 4). The planar form of *a6* is lying at 'only' 3.5 eV above the bowl form, whereas, as mentioned, the planar form of *f6* is lying as high as 5.2 eV above the bowl form, again emphasizing how fulvene inclusions along a PAH rim brings much larger strains than azulene inclusions.

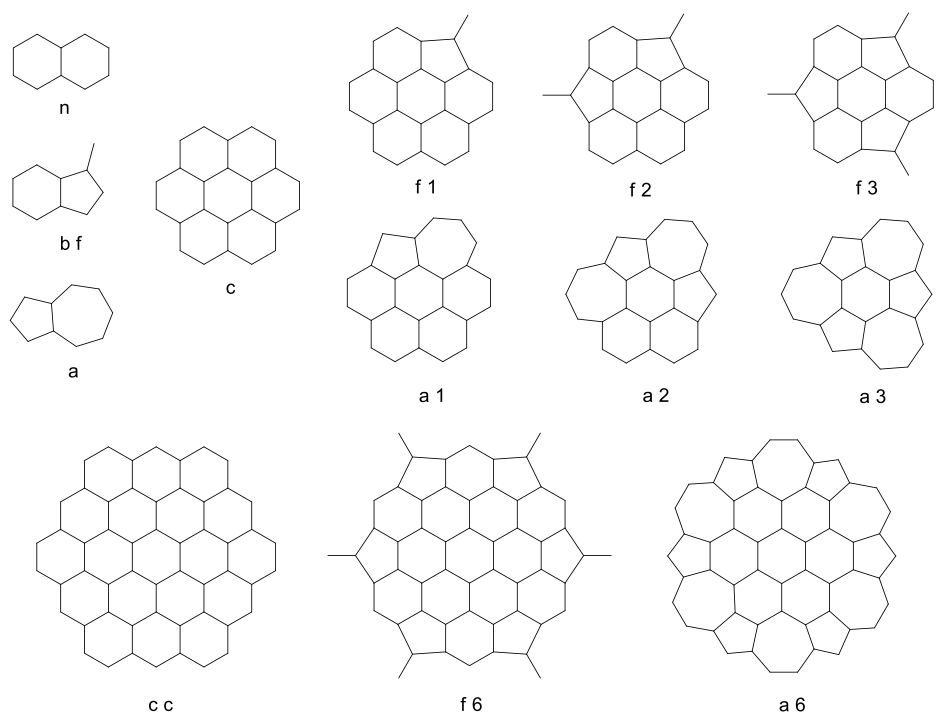

Scheme 10

Strict delocalization effects obtained from HMO treatments are not as clear as what is suggested above. In the coronene series, they would favor *f1*, *f2* and *f3* over their corresponding azulene counterpart (Table 4, right column). DFT-calculated energies actually favor *a2* and *a3* over *f2* and *f3*, but this can be accounted for from strain arguments. These no longer holds for naphthalene isomers, where, as seen above, DFT gives benzo-fulvene below azulene, contrary to HMO results.

For coronene and circum-coronene series, Table 4 suggests a reasonable transferability of energy increments needed to azulene substitution, although to a lesser degree than with vinylidene bridges. To some extent, HMO energies also reflect transferability, although somewhat unequally: excellent in *a1* to *a3*, poorer in *a6* or *f6*. Discussing such imprecisions or discrepancies, one should keep in mind that, *stricto sensu*, the use of Clar's sextets concerns 'regular' electron sextets resting upon *neutral* carbon six-membered rings, so that Clar reasoning on sextets resting upon *formally charged* five- or seven-membered rings may not be that valid. In any case, the barrier toward azulene-type isomers should be higher than those associated to vinylidene formation. In the former route, there are two carbon-carbon bonds to be broken and reassembled, implying four carbon atoms, two electron doublets and conjugation loss at two adjacent cycles, while vinylidene formation, if implying additional 1→2 hydrogen shift, involves only one carbon-carbon bond to be broken and reassembled, a single electron doublet migration and conjugation breaking at only one ring.

**5. Twisted forms**.

We now address a rather atypical category of high-lying local minima, in which one or several double bonds adopt a twisted configuration, with H-C-C-H in *trans* arrangement (Scheme 11). To understand their origin, one must trace back to *cis-trans* isomerization of secondary olefins, as detailed in Appendix E, which also addresses the computational methods to reliably describe their energies and structures. To begin with, the existence of such peculiar arrangements is tested on the basic cyclic olefins. In cyclopentene, no twisted minimum is found at CAS(2,2) level, probably due to the huge strain it would require. Such a minimum is however found at DFT level, which we might not trust here. In cyclohexene, both levels found the twisted minimum as a local minimum lying a little more than 2 eV above the regular $C_s$ form close to planar. In benzene, DFT, CAS(2,2), CAS(4,4) and CAS(6,6) found the twisted minimum, here lying more than 4 eV above the planar regular ground state, twice as high as in cyclohexene, as expected from the breaking of aromatic resonance. In cyclohexene, the barrier required to return to the regular form is at 0.79 eV above the twisted form (CAS(2,2) level), reduced to 0.68 eV after ZPE correction, a reasonable and significant barrier of 16-18 kcal/mol. Not surprisingly, that to be overcome for returning to normal benzene is rather small, although it seems to persist after ZPE corrections. At CAS levels, the barrier is estimated at only 0.12-0.14 eV (raw), reduced to 0.06-0.08 eV (ZPE). So tiny values might here question the twisted-minimum existence, as it could vanish at higher levels of treatment.

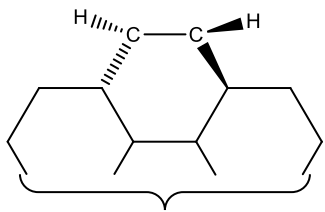

Scheme 11

The twisted forms of coronene, pyrene and phenanthrene that have been explored all lie at 3.3-3.7 eV above normal planar forms (Table 5), with return barriers around 0.2 eV (5 kcal/mol). This is a weak value, making this forms likely to be visited at high temperature, but unlikely to be present after cooling of the system. Tipping over an outer –CH=CH- edge into a twisted high-lying minimum therefore seems ubiquitous in PAHs. These tilted forms share common features for the C–HC=CH–C skewed moiety: perfect *trans* arrangement for H–C–C–H, C–C–C–C dihedral angle ω of 76°, twisted-bond length $d_{c-c}$ of 1.40 Å. This also holds for the transition states leading to the twisted minima: C–C–C–C dihedral angle ω of 59°, twisted-bond length $d_{c-c}$ of 1.47 Å.[1] An illustration of the twisted geometry is given in Figure 6, for neutral coronene and pyrene. Again, it is possible to wring several HC=CH edges along the outer rim of a PAH, with a transferability gauged in the bottom of Table 5 (further discussion on this point is given in the supplementary material S13). In conclusion, this kind of high-lying alteration of the outer edges is expected to be ubiquitous in PAHs. Strictly, it does not change the topology of a given PAH, but it can be seen as a first step toward, or anticipating, another kind of isomers, bearing bicyclobutane parts, and addressed in the following section.

**6. Bicyclobutane forms.**

Further tilting the above-discussed skewed HC=CH groups to full orthogonal arrangement leads to local bicyclobutane motifs (Scheme 12). In these forms the orthogonal H–C–C–H set is now in *cis* position, and conjugation patterns have been dramatically modified, diminishing the number of Clar's sextets, or even destroy them in pyrene. These forms are therefore expected to lie rather high above the normal planar forms. They correspond to $C_{2v}$ geometries, with the main skeleton remaining planar, and their relative energies are given, referenced as *b-1*, in Table 6. They stand at more than 5 eV for neutral coronene and pyrene, and at 3-4 eV in their cationic states. In phenanthrene, the form is about 1-eV lower, probably due to less angular strains. These isomers can be attained from the twisted forms, or directly from the normal forms. In any cases, return barriers are expected to be weak, as we have checked in a few instances. Illustrations given in Appendix F compare the indirect normal-to-bicyclobutane pathway *via* a twisted form for coronene dication with the direct normal-to bicyclobutane pathway for coronene trication. In both cases, the calculated return barriers from bicyclobutane forms are weak (0.1-0.2 eV), but they persist to ZPE corrections.

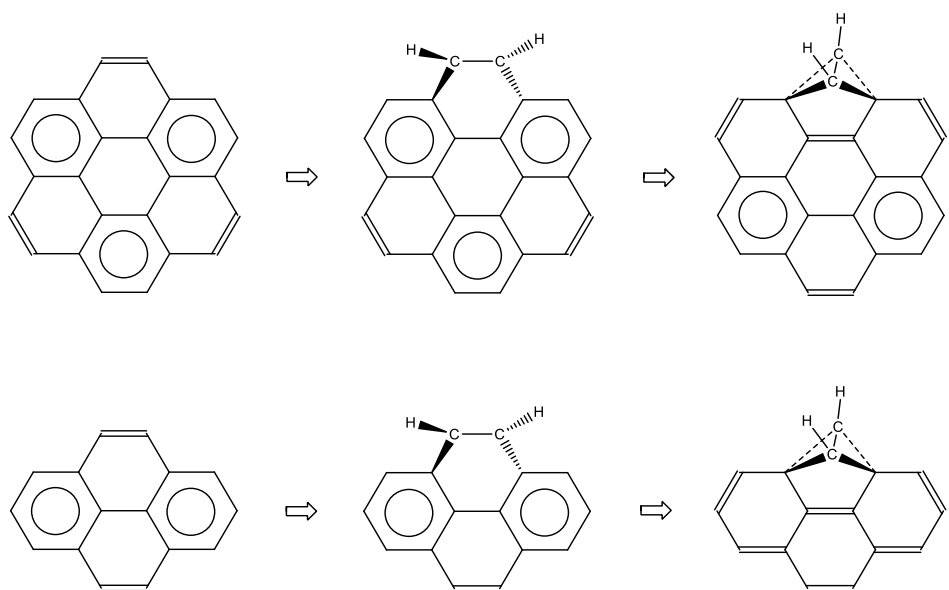

Scheme 12

Actually, there exists a less symmetrical and less obvious bicyclobutane form that competes with the *b-1* $C_{2v}$ form. At first glance, the arrangements *b-2* in Scheme 13 are less symmetrical, and their bicyclo branching should (and does) induce loss of planarity at the conjugated carbon framework. However, more attentive inspection indicates that conjugation is here more important than in the *b-1* arrangement, as illustrated in the graphs given in Appendix G, and made more quantitative in the HMO results given in the supplementary material section (Table S14). One can also notice in the top of Scheme 13 that there exist two degenerate sets of pairs of Clar's sextets in *b-2*, whereas only one set is present in *b-1*. For neutral coronene, *b-2* is found to be more stable than the *b-1* by 0.36 eV (ZPE, 0.32 eV). Actually, this alternative bicyclo form is no longer preferred in the coronene monocation, and even ceases to exist in coronene dication and trication (see Table 6, bottom line).

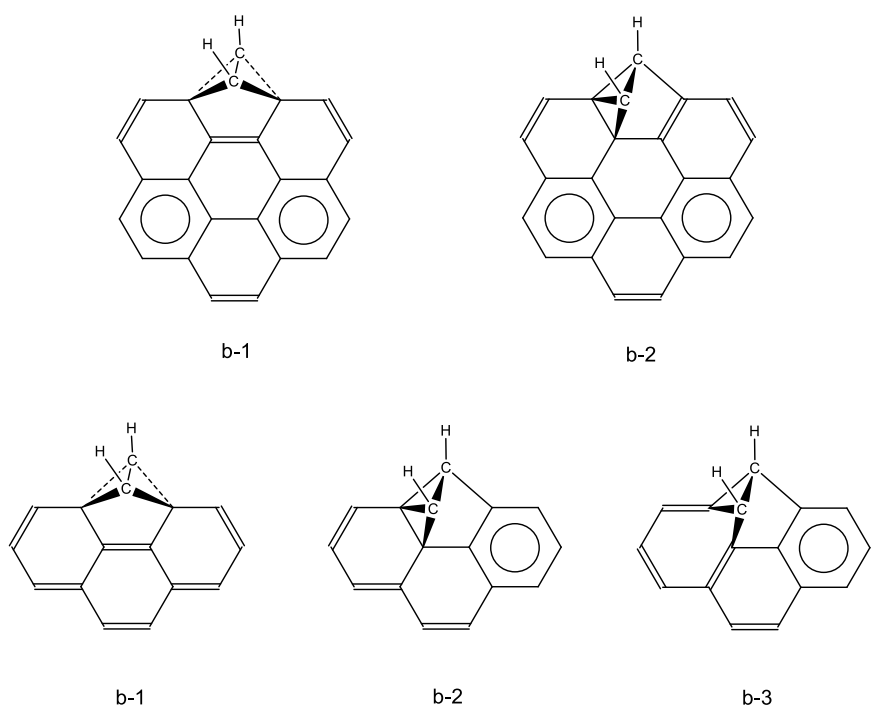

Scheme 13

In neutral pyrene, *b-2* is lying 0.67 eV below *b-1*, but in this case it only corresponds to a plateau point, eventually caught in a lower minimum no longer comprising the bicyclobutane motif and labelled *b-3* in Scheme 13. By contrast, in all cationic pyrenes, *b-2* is found to be a real minimum, but only for the monocation is *b-2* more stable than *b-1*, as detailed in Table 6. For phenanthrene, *b-2* exists as a true minimum only in the neutral state, where it is lying 0.73 eV below *b-1*. Such features are not surprising, given that existence and relative positions of these minima are affected by frame rigidity, particularly weak in altered phenanthrene.

A transition state linking twisted and *b-2* forms has been found for neutral coronene. From this bicyclobutane form, a barrier of 0.84 eV (ZPE, 0.76 eV) is now required to reach the twisted form, lying 1.2 eV below. Along this path, *b-2* is thus efficiently trapped. By contrast, in the coronene cation, the barrier required to return from *b-2* to the normal form, lying 4.6 eV below, is found at only 0.10 eV (ZPE, 0.07 eV), questioning the viability of this secondary minimum. A similar observation can be made for the pyrene cation from the energies recapitulated in Appendix H. For pyrene dication, a transition state linking *b-1* and *b-2* (Scheme 14) has been found, corresponding to a barrier of 0.59 eV above *b-1* (ZPE, 0.55 eV), and 0.29 eV above *b-2* (ZPE, 0.22 eV).

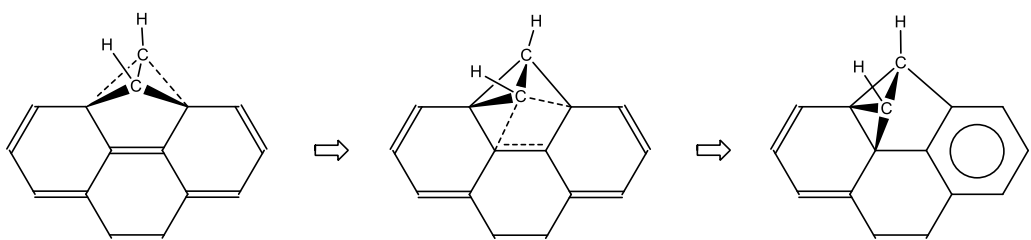

Scheme 14

As mentioned above, in neutral pyrene, *b-2* relaxes into *b-3* by breaking a bicyclobutane C–C bond. Because of this bond breaking, *b-3* is in fact more conjugated than *b-2*, as illustrated in Appendix G (and made more quantitative in Table S14). As such, it can be expected to be more stable as long as one keeps with delocalization effects and disregard stereochemical constraints. Actually, this is the case in several instances, as can be seen in Table 6. The *b-3* structure is catching the *b-2* one for neutral pyrene and cationic phenanthrene, while it is the reverse for tri-cationic pyrene. For coronene, this form is always a real minimum, but never a catchment region from *b-2*. In contrast to pyrene and phenanthrene, *b-3* is here above *b-2* for the neutral form. The full HMO analysis of deconjugation due to bicyclo architectures is given in Table S14 of the supplementary material.

Some barriers have been calculated to escape from *b-3* (see Appendix H). Towards the twisted form, a barrier of 0.45 eV (ZPE, 0.39 eV) is required for coronene di-cation and a barrier of 1.26 eV (ZPE, 1.18 eV) is required in neutral pyrene. Towards the normal form, a tiny barrier of only 0.10 eV (ZPE, 0.06 eV) is calculated in coronene trication. Again, it is possible to dispose several bicyclobutane frames around large PAHs. From the relative energies of bis- and tris-bicyclobutane forms of coronene given the bottom of Table 6, decent transferability is deduced (see more comments in the supplementary material S15). Finally, some optimized geometries are also given as supplementary material (Figure S16), illustrating the extent of nonplanar deformation possibly brought by the bicyclobutane branching.

**7. Interconnections and dissociations.**

As seen in the first companion paper, formation of $CH_2$ motif through hydrogen shift may lead to broad hydrogen exchange (see Appendix I, right). Carbon atom scrambling is also possible *via* vinylidene or bicyclobutane intermediates. Strictly speaking, the "direct" mechanism for vinylidene formation would switch carbons but not hydrogens (see Appendix I, left), were it not for the possibility of torsion around extracyclic $C=CH_2$ in vinylidene minima. Such double-bond isomerization, which costs 2-3 eV, should allow the scrambling of both carbon and hydrogen within a –CH–CH– edge, as does the indirect mechanism visiting hydrogen-shifted minima (see Appendix I, right). While bicyclobutane intermediates can switch CH groups within a –CH–CH– edge, ethynyl alternatives do not bring carbon scrambling, except in those cases which admit some free rotation between phenyl groups in pyrene or phenanthrene (structures *f*, *g* and *h* in Figure 1).

Scheme 15 summarizes the connections between the various PAH isomers addressed in the present work. Each form can eliminate specific fragments, and we have explored a few cases of them, namely the elimination of H, $C_2H_2$ and $C_2H_4$ from normal forms, and that of $H_2$ from hydrogen-shifted isomers. Asymptotic adiabatic dissociation energies for such eliminations are given in Table 7.

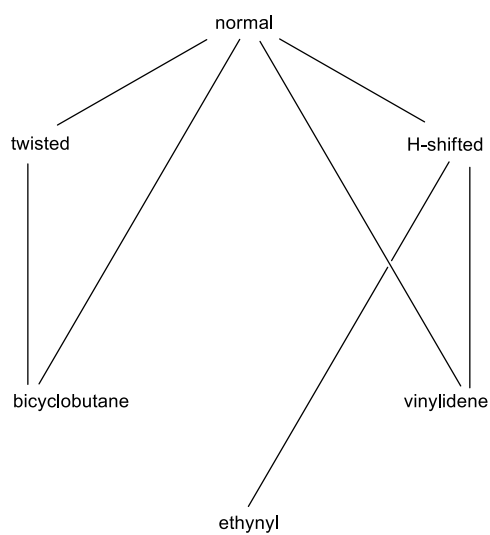

Scheme 15

Atomic hydrogen elimination proceeds through simple homolytic cleavage. While there exists only one kind of single hydrogen removal in coronene, there are three possibilities for pyrene, as shown in Scheme 16, top. In all neutral species, this dissociation energy is calculated at 5.1 eV (ZPE, 4.7 eV), whatever the site of hydrogen loss in pyrene. In coronene, the dissociation energy for cationic species is larger, except for dication which is similar to the neutral form. In pyrene, this dissociation energy is also larger for cationic states, except for trication, where it is lower. In addition, as already revealed in the work by Chen *et al.* (2015), we have here a nice site-specificity for the monocation. If dehydrogenation occurs at the tip of the pyrene structure (position labelled *5*), it makes it possible to delocalize the positive charge in a phenalene cation, while organizing an allene part around the hydrogen-depleted site, as illustrated in Scheme 17. This feature, distinctive of PAH including a 'protruding' six-membered ring, is plainly confirmed by calculated CC bond lengths, non-planar geometry, and electronic charges. As a consequence, and as noticed previously (Chen et al. 2015), the dissociation energy is here calculated at 5.1 eV, while at positions *1* and *4*, as well as in mono-cationic coronene, it is calculated at 5.3 eV (ZPE values, see Table 7).[2,3] When the system contains enough energy to dissociate, these various channels, with minor energy differences, are likely to compete.

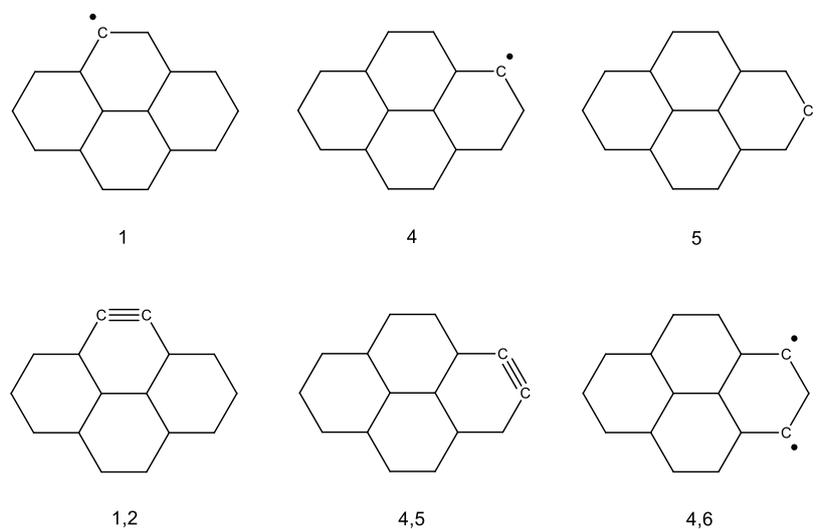

Scheme 16

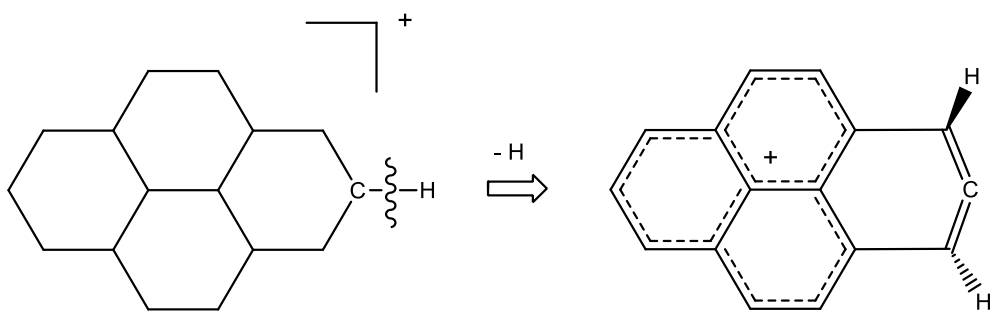

Scheme 17

Elimination of two hydrogens, either as separated atoms or under $H_2$ molecular form, again gives rise, in pyrene, to several topomers (Scheme 16, bottom). Besides the two vicinal dehydrogenated forms 1,2 and 4,5, the non-vicinal form 4,6 is to be taken into account as it may further cyclize into a cyclopropenyl motif, which is a stabilizing factor in cationic species. Putting aside this pyrene (4,6) case, the adiabatic energies for elimination of H+H or $H_2$ are close in pyrene and in coronene, with a regular increase as a function of the positive charge, namely from 4 to 5 eV for $H_2$ elimination, and from 8 to 9 eV for H+H elimination (ZPE level, cf Table 7). In contrast, for the (4,6) alternative of pyrene, the dissociation energy is high in the neutral form, while it is the lowest one in the dication. In any case, this arrangement binds the carbon centers 4 and 6 in a cyclopropenyl three-membered ring. When positively charged, this motif is endowed with aromatic character, and one understands that the full structure is particularly favorable to dicationic configurations, where the two charges can delocalize within distinct parts of the carbon skeleton, namely a phenalene cation and a cyclopropenyl cation, as illustrated in Scheme 18 (for aromaticity relocation in normal pyrene dication, see Matsumoto et al. 2014). Again, this view is confirmed by the calculated geometries and charges, and it also holds, to a lesser extent, for tricationic pyrene. In this way, the dissociation energy of dicationic pyrene for $H_2$ elimination is as low as the neutral coronene and pyrene values, around 3.8 eV (ZPE level).

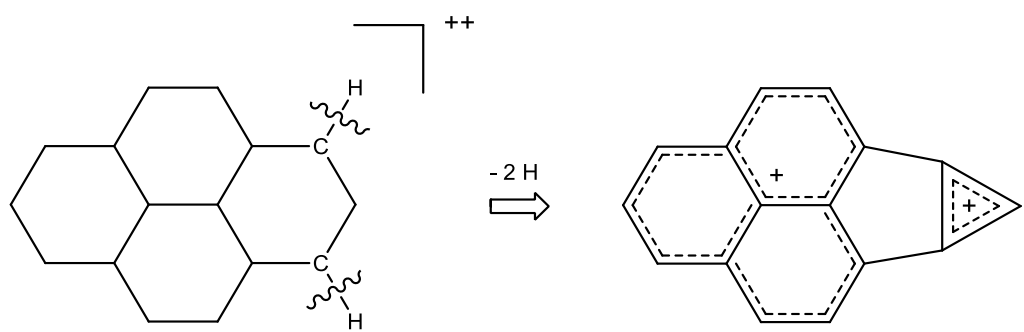

Scheme 18

The lower path for concerted $H_2$ elimination proceeds *via* hydrogen-shifted isomers (Chen, et al. 2015; Jolibois et al. 2005). The barriers to overcome from 1,2-hydrogen shifted isomers are reported in Table 8 (overall from normal forms), with corresponding energy profiles plotted in Figure

7. In neutral species, these barriers are calculated at 1.6-1.7 eV, and in cationic species, they amount to 2.3-2.7 eV, with a low value of 1.9 eV for pyrene dication. The overall barriers calculated from normal forms are all around 5 eV. For pyrene dication, proceeding *via* the 5,6-hydrogen shifted isomer is overall more advantageous. The elimination barrier from this minimum is larger than that from the 1,2 type, but this is over-compensated by its energy, significantly lower than that of the 1,2-type, so that finally the overall barrier via the 5,6-hydrogen shifted isomer (4.8 eV) is lower than that *via* the 1,2 isomer (5.0 eV), as found previously (Chen, et al. 2015). According to the results of Tables 7 and 8, it is therefore easier to eliminate $H_2$ than H in coronene monocation and trication, and in pyrene monocation and dication.

Let us now address the elimination of larger fragments like acetylene or ethylene, in a simplified illustrative approach, limiting to direct and symmetrical emblematic sequences. We notice, first, that when the starting material is under multicationic state, the process is expected to be facilitated if the eliminated fragment takes away a positive charge. In these cases, dissociation limits are expected to be lower than in the cases where the remaining fragment retains the whole multicationic charge. As can be checked in the bottom part of Table 7, this is actually observed, except for acetylene elimination from dicationic coronene. Because ionization energy of acetylene, here calculated at 11.26 eV, is higher than that of the remaining fragment $C_{22}H_{10}^+$, here calculated at 10.57 eV, it is more advantageous, adiabatically, to dissociate $C_{24}H_{12}^{++}$ into ($C_{22}H_{10}^{++} + C_2H_2$) than to dissociate it into ($C_{22}H_{10}^+ + C_2H_2^+$), not to mention the barrier inherent to dissociation reactions with charge separation.

Ionization energies of all fragments involved in acetylene and ethylene eliminations are plotted in Figure 8, suggesting the following more-or-less obvious trends. A starting monocation is expected to always favor elimination of neutral small organic fragments, while a starting trication is expected to always favor elimination of cationic small organic fragments. When the starting PAH is a dication, from coronene to larger PAHs, elimination of neutral acetylene is preferred over cationic acetylene loss. Regarding elimination of ethylene, the loss of cationic ethylene is favored in the case of pyrene and coronene dications, while elimination of neutral ethylene is likely for larger PAHs dications.

Exploring direct elimination of acetylene from convex –HC–CH– sets shows that both transition state and remaining fragment bear a four-membered ring, inducing bowl-shaped geometries (Scheme 19).[4] In neutral coronene, the elimination proceeds in one-step concerted way, while in the monocation, there exists a high-lying local minimum, as illustrated by Figures 9 and 10. Such high-lying secondary minimum intermediates seem to exist also for coronene dication, and pyrene monocation and dication. Changing density functional or basis set might modulate their existence in the course of these elimination paths, but this should not change overall barriers. For trications, the mechanism seems more complex and we could not get a clear picture of it; similarly, we fail to obtain the transition state for acetylene elimination from neutral pyrene. Finally, overall barriers are calculated at 9.1, 8.1 and 7.2 eV for neutral, monocationic and dicationic coronene, respectively (ZPE, 8.9, 7.9, 7.1 eV, see Table 9), and at 7.6 eV and 7.2 eV for monocationic and dicationic pyrene respectively (ZPE, 7.4, 6.9 eV), as summarized in Table 9.

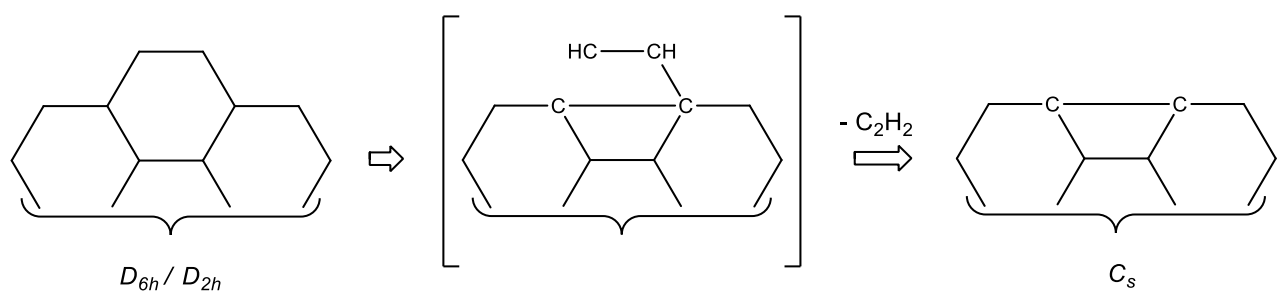

Scheme 19

For pyrene, taking out a $C_2H_2$ fragment from the "lateral" ring may appear more advantageous, as revealed in the computational and experimental explorations of unimolecular dissociation of pyrene radical cation by West et al. (2014). In such fragmentation (type 2 in Scheme 20), the resulting four-membered ring is better inserted in the structure than it was in the former type-1 case, inducing less strain, hence less puckering, hence less conjugation loss. Actually, such $C_{14}H_8$ fragment is found to be only slightly warped in its neutral and monocationic forms, and fully planar in its di- and tricationic forms. As indicated in Table 7, for the monocationic state, the dissociation asymptote for this type of fragmentation is calculated at 6.0 eV (ZPE level), a value which is in agreement with that of West, et al. (2014), and lower by 0.8 eV than that associated to the former type-1 case. For the neutral state, the type-2 process is no longer favored, with a dissociation energy now 0.4-eV higher than that of type 1. This feature may be accounted for from the strong bond alternation of the *neutral* cyclobutadiene part,[5] inducing some loss of delocalization - formally, we have here only one Clar's sextet (Scheme 21, top). By contrast, for the dicationic state, type-2 fragmentation is favored over type-1 by as much as 1.5 eV, a result that can be accounted for by noting that the cyclobutadiene part can now take a dicationic aromatic form. Such formal writing does not optimize charge distribution, but it restores two Clar's sextets (Scheme 21, bottom). This naïve view of the electronic structure is supported by the calculated planar geometry and Mulliken charges. Not surprisingly, for the radical cation, both charges and spin densities favor a scheme (phenalene cation + methyl radical) over its symmetrical counterpart (Scheme 21, middle). The above-discussed relative stability of fragment $C_{14}H_8^{++}$ in its type-2 form entails another consequence, namely, that the elimination of cationic acetylene from tricationic pyrene ($C_{16}H_{10}^{+++} \rightarrow C_{14}H_8^{++} + C_2H_2^+$) has now an asymptotic dissociation energy as low as 0.3 eV (ZPE level; see Table 7, bottom).

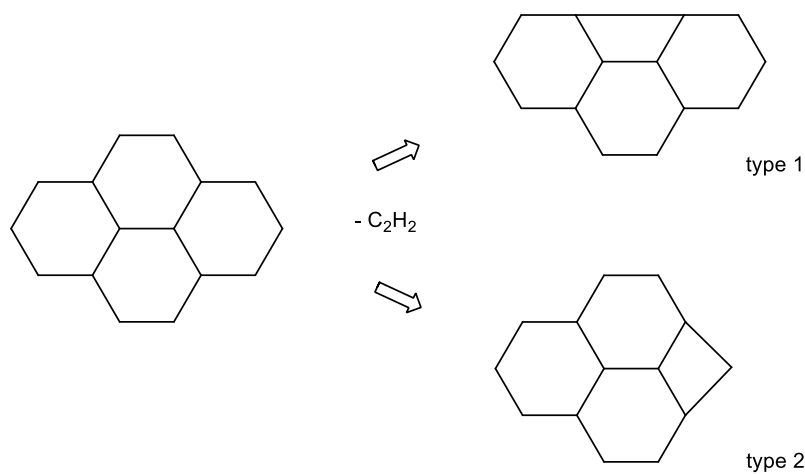

Scheme 20

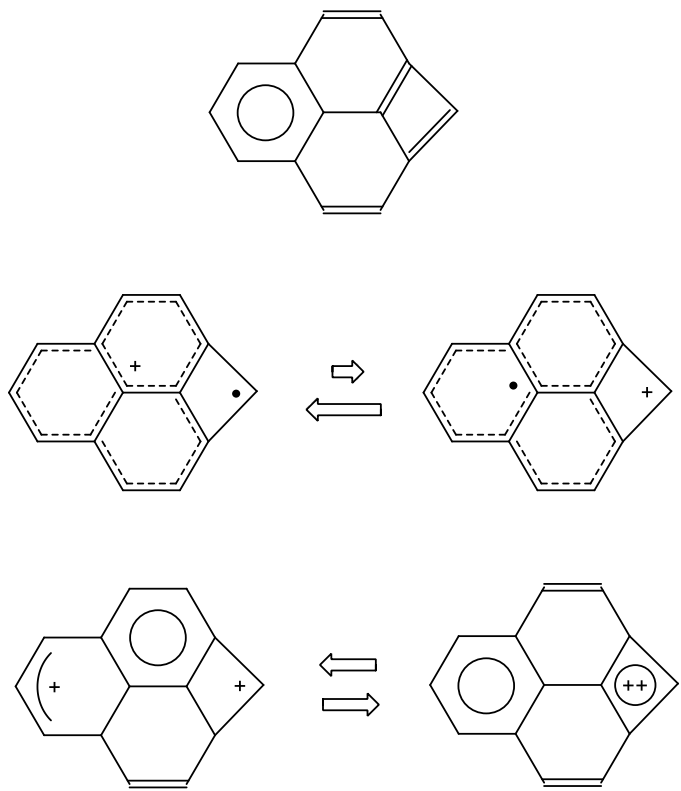

Scheme 21

We address, last, the possible direct elimination of ethylene from a 'prepared' doubly hydrogen-shifted isomer (Scheme 22). With respect to normal forms, these starting structures are lying at 4.8-5.3 eV for coronene (ZPE, 4.7-5.2 eV), and 4.5-6.6 eV for pyrene (ZPE 4.3-6.5 eV, see Table 10), with additional barriers around one eV to reach them from the normal forms. Transition states corresponding to this concerted elimination have been found for neutral and monocationic species (their structures are given in the supplementary material (Figure S17). In both neutral forms, these dissociation barriers are lying at 0.6 eV above the dissociation limit. In cationic coronene, this value is reduced to 0.3 eV, while in cationic pyrene, the barrier disappears, the number provided in Table 10 corresponding here to the barrier separating the starting system $C_{16}H_{10}^+$ from the intermolecular complex ($C_2H_4 + C_{14}H_6^+$). For multications, the barriers no longer correspond to direct concerted pathways. They are higher above the dissociation limits, since these are lowered as they involve cationic ethylene. Overall barriers, reported in Table 10, are around 8.8 eV.

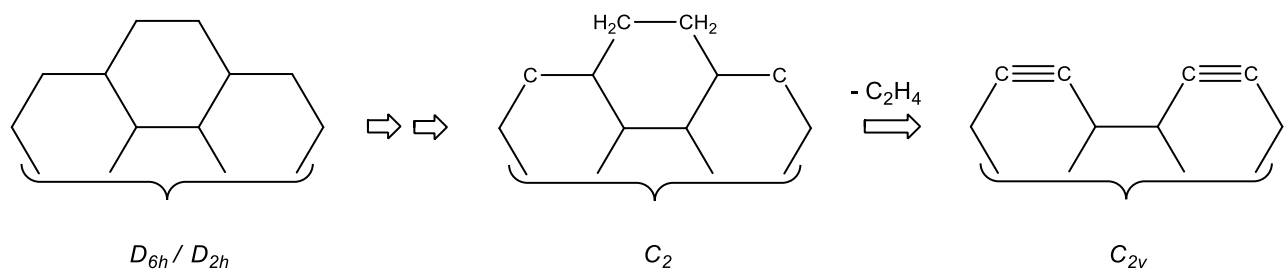

Scheme 22

## 8. Conclusion.

Dealing with a limited sample of possible rearrangements mainly examined on two emblematic representatives of PAH systems, this survey nevertheless provides some hints on potential evolution of such species, in their ground-state neutral or cationic forms, as soon as they carry substantial amounts of internal energy (say up to 5-10 eV). Within these ranges of internal energies, carbon and hydrogen atoms are entitled to occupy various positions on the potential-energy surfaces, with some corresponding arrangements leading to different dissociation processes. Such alternatives structures are more or less stable, both kinetically or thermodynamically, and as such they may play central parts in PAH dynamical processes. Among these, we found the familiar arrangements of organic functions, from classical ones such as triple bonds (opened rings) or benzofulvenes (vinylidene alternatives), to less common ones such as methylcarbenes (hydrogen-shifted isomers) or bicyclobutanes, or even more exotic ones like these "twisted forms" that seem to be endemic to PAH frames. The traditional concepts of organic chemistry therefore remain valid at several electron-volts of internal energy, as remain valuable some of its classical tools such as Clar's rules or Woodward-Hoffmann's rules, that can be put to profit to understand structures and guess thermodynamic or kinetic stabilities.

The presently-studied alternative structural isomers are expected to play significant parts in various instances where PAHs are brought to such energies, as in collisions with protons or electrons, or upon absorption of UV radiations, and are expected to have incidence on vibrational signatures and on the production of fragments of various sizes. Among significant features documented in the two parts of this work, we highlight (1) the possible « walk » along the skeletal rim of both hydrogen atoms on saturated sites, and holes on hydrogen-deficient sites (i.e. carbenic centers), (2) the relatively easy routes to the vinylidene (or benzofulvene) forms which now affect carbon backbones, (3) the relatively low-lying ethynyl and 2-butynyl forms, resulting from ring openings. Formation of such arrangements should be ubiquitous in PAHs and are to be taken into consideration whenever energy has been deposited on them, not only in the interstellar medium when submitted to proton bombardment and UV photons, but also in combustion, ionization, and collision laboratory experiments. It is even possible that they can be trapped into these higher-energy isomers following infrared cooling in the isolated conditions of the interstellar medium. These structures should imprint the dynamics and processing of interstellar PAHs, regarding their role as precursors of smaller species like H, $H_2$, $C_2H_2$. Furthermore, the easy and various paths to pentagon formation underlined in this work could open interesting routes to fullerene formation investigations.

As PAH size increases, so does the number of secondary minima, and their interconversion may involve intricate processes. Amongst various approaches, a convenient tool for grasping these competitions and complexities is to perform molecular dynamics. Actually, some of the presently-described alternative structures have been seen as elusive intermediates in recent studies of molecular dynamics of PAH dissociation (Simon et al. 2017). In summary, while PAHs remain relatively rigid at low energy, when they carry several electron-volts of internal energy, they become ductile and exhibit more or less plasticity, which should guide their evolution before and during fragmentation - energized PAHs are not PAHs at rest.

**Acknowledgments** This work is part of the SWEET project (Stellar Wind and Electron interactions on astrophysical molecules. Experiment and Theory), integrated in the NEXT collaborative project, and involving teams of LCAR (CNRS-UMR5589), LPT (CNRS-UMR5152), and LCPQ (CNRS-UMR5626) laboratories.

**Notes**

(1) Averaged over sixteen CAS calculations on coronene, pyrene and phenanthrene, the mean values and standard deviations for $\omega$ and $d_{c-c}$ are: 75.9° (1.2°), 1.398 Å (0.003 Å) for twisted minima, and 58.5° (1.6°) and 1.467 Å (0.003 Å) for transition states.

(2) In Chen et al. 2015, it seems labels *a* and *b* have been switched in Figure 4, left, set ($C_{16}H_9^+$ + H).

(3) Our mono-dehydrogenated cationic fragments, bearing an even number of electrons, actually have a triplet ground state, located at about 0.5 eV below the closed-shell singlet (except for the case of pyrene position 4, where this gap is only 0.2 eV). The true asymptotes would therefore here correspond to triplet states, lying below the singlet ones by these differences. In naphthyl cation $C_{10}H_7^+$, this singlet/triplet competition is particularly tight (Dutta et al. 2014). Similarly, the ground state for mono-dehydrogenated pyrene trication (position 1) if found to be a triplet, with the closed-shell singlet lying above by 0.6 eV.

(4) For some dissociated fragments, $C_{2v}$ planar forms are found to be real minimum, lying at 0.5-1.0 eV above the preferred non-planar minima.

(5) The CC bond lengths in the cyclobutadiene motif are calculated as follows (Å). Neutral fragment: 1.38, 1.63 (outer), 1.37, 1.45 (inner); monocation: 1.50 (outer), 1.40 (inner); dication: 1.47 (outer), 1.41 (inner); trication: 1.48, 1.49 (outer), 1.39, 1.43 (inner).

Table 1. Relative energies for isomers resulting from ring breaking and including triple bonds.[a]

| | | | | | ΔE (eV) | |
|---|---|---|---|---|---|---|
| | | | | | Raw | ZPE |
| coronene | neutral | ethynyl | $C_1$ | a | 2.71 | 2.60 |
| | | di-ethynyl | $C_1$ | b | 5.33 | 5.12 |
| | | | $C_2$ | c | 5.35 | 5.14 |
| | | tri-ethynyl | $C_1$ | d | 7.89 | 7.58 |
| | | TS(1,2-ethynyl) | $C_1$ | | 4.67 | 4.54 |
| | monocation | ethynyl | $C_1$ | a | 2.48 | 2.44 |
| | dication | ethynyl | $C_1$ | a | 2.19 | 2.13 |
| | trication | ethynyl | $C_1$ | a | 2.33 | 2.31 |
| pyrene | neutral | ethynyl | $C_S$ | e | 2.47 | 2.36 |
| | | di-ethynyl | $C_2$ | f | 4.64 | 4.42 |
| | | | $C_2$ | g | 4.60 | 4.38 |
| | | TS(1,4-ethynyl) | $C_1$ | | 5.42 | 5.22 |
| | monocation | ethynyl | $C_S$ | e | 2.83 | 2.71 |
| | dication | ethynyl | $C_S$ | e | 3.06 | 2.93 |
| | trication | ethynyl | $C_1$ | e | 2.47 | 2.37 |
| phenanthrene | neutral | ethynyl | $C_1$ | h | 2.28 | 2.17 |
| coronene | neutral | 2-butynyl | $C_1$ | i | 4.07 | 3.99 |
| | | di-2-butynyl | $C_2$ | j | 8.24 | 8.07 |
| | | | $C_1$ | k | 8.26 | 8.09 |
| | | TS(1,19-2-butynyl) | $C_1$ | | 4.79 | 4.67 |
| pyrene | neutral | 2-butynyl | $C_1$ | l | 3.78 | 3.71 |
| | | | $C_1$ | m | 3.95 | 3.86 |
| | | di-2-butynyl | $C_1$ | n | 6.98 | 6.79 |
| | | | $C_1$ | o | 5.56 | 5.39 |
| | | TS(4,16-2-butynyl) | $C_1$ | | 4.69 | 4.56 |
| phenanthrene | neutral | 2-butynyl | $C_1$ | p | 3.75 | 3.66 |
| | | | $C_1$ | q | 3.77 | 3.66 |
| | | | $C_1$ | r | 3.67 | 3.58 |

[a] See Figures 1 and 2 for labeling.

Table 2. Relative energies (eV) of various vinylidene forms with respect to normal forms.[a]

| | | | raw | | | | ZPE | | | |
|---|---|---|---|---|---|---|---|---|---|---|
| | | | neutral | monocation | dication | trication | neutral | monocation | dication | trication |
| coronene | mono-vinylidène | $C_{2v}$ | 1.60 | 1.30 | 1.06 | 1.14 | 1.55 | 1.30 | 1.04 | 1.17 |
| | di-vinylidène | $C_s$ | 3.58 | 3.36 | 3.11 | 3.15 | 3.48 | 3.32 | 3.06 | 3.15 |
| | tri-vinylidene | $C_{3v}$ | 5.40 | 5.45 | 5.43 | 5.34 | 5.26 | 5.32 | 5.31 | 5.24 |
| pyrene | mono-vinylidène | $C_{2v}$ | 1.16 | 1.45 | 1.67 | 1.17 | 1.11 | 1.38 | 1.61 | 1.13 |
| | di-vinylidène | $D_{2h}$ | 3.34 | 3.56 | 3.90 | 3.30 | 3.23 | 3.44 | 3.77 | 3.19 |
| phenanthrene | mono-vinylidène | $C_{2v}$ | 0.69 | 0.74 | 0.91 | 0.71 | 0.65 | 0.69 | 0.82 | 0.66 |

[a] Symmetries refer to neutral species.

Table 3. Energy barriers (eV) to be overcome for reaching mono-vinylidene isomers.

| starting point | charge state | raw | | ZPE | |
|---|---|---|---|---|---|
| | | coronene | pyrene | coronene | pyrene |
| normal form | neutral | 5.34 | 4.91 | 5.17 | 4.75 |
| | monocation | 5.07 | 5.20 | 4.91 | 4.96 |
| | dication | 4.96 | 5.75 | 4.78 | 5.49 |
| | trication | 4.52 | 4.28 | 4.39 | 4.14 |
| 1,2-H-shifted isomer | neutral | 4.40 | 4.22 | 4.30 | 4.12 |
| | monocation | 3.90 | 3.79 | 3.85 | 3.69 |
| | dication | 3.63 | 3.66 | 3.55 | 3.55 |
| | trication | 3.42 | 3.14 | 3.38 | 3.00 |

Table 4. Relative energies (eV) of fulvene-type and azulene-type alternatives with respect to normal naphthalene, coronene, and circum-coronene (see Scheme 10 for structure definitions).

| reference PAH | alternative | raw | ZPE | HMO (|β|) |
|---|---|---|---|---|
| naphthalene | azulene | 1.47 | 1.43 | 0.32 |
| | benzo-fulvene | 1.05 | 1.00 | 0.35 |
| coronene | *a 1* | 1.78 | 1.74 | 0.33 |
| | *f 1* | 1.60 | 1.14 | 0.28 |
| | *a 2* | 3.28 | 3.21 | 0.66 |
| | *f 2* | 3.58 | 3.15 | 0.53 |
| | *a 3* | 4.44 | 4.32 | 0.98 |
| | *f 3* | 5.40 | 5.34 | 0.76 |
| circum-coronene | *a 6* | 9.82 | 9.71 | 1.34 |
| | *f 6* | 9.81 | 9.63 | 1.34 |

Table 5. Calculated relative energies for some twisted forms (in eV).

|  | DFT | CAS |
|---|---|---|
| 2-butene [a] | 2.24 | 2.17 |
| cyclopentene [b] | 4.07 |  |
| cyclohexene | 2.35 | 2.39 |
| benzene | 4.64 | 4.27 |
| phenanthrene | 3.64 | 3.35 |
| pyrene | 3.68 | 3.27 |
| coronene | 3.94 | 3.65 |
| coronene dication | 3.48 | 3.29 |
| pyrene bi-twist | 7.35 | 6.63 |
| coronene bi-twist | 7.65 | 6.91 |
| coronene tri-twist | 11.32 | 10.29 |

[a] CCCC dihedral angle constrained to 75°; relative energy with respect to regular *trans* form.

[b] No twisted minimum found at CAS level.

Table 6. Relative energies (eV) for structural isomers with bicyclobutane parts. [a]

| system | minima | raw | | | | ZPE | | | |
|---|---|---|---|---|---|---|---|---|---|
| | | 0 | + | ++ | +++ | 0 | + | ++ | +++ |
| coronene | *b-1* | 5.27 | 4.47 | 3.75 | 3.94 | 5.15 | 4.42 | 3.68 | 3.88 |
| | *b-2* [b] | 4.91 | 5.58 | - | - | 4.83 | 4.54 | - | - |
| | *b-3* | 5.08 | 4.48 | 3.94 | 4.27 | 5.00 | 4.46 | 3.93 | 4.27 |
| pyrene | *b-1* | 5.39 | 4.50 | 3.75 | 3.37 | 5.25 | 4.37 | 3.64 | 3.27 |
| | *b-2* [c] | 4.72 | 4.41 | 4.05 | 3.59 | - | 4.33 | 3.97 | 3.52 |
| | *b-3* [d] | 4.35 | 4.40 | 4.45 | 4.08 | 4.28 | 4.31 | 4.33 | - |
| phenanthrene | *b-1* | 4.64 | 3.36 | 2.16 | 2.31 | 4.51 | 3.27 | 2.10 | 2.24 |
| | *b-2* [e] | 3.91 | 3.69 | 3.27 | 3.35 | 3.84 | - | - | - |
| | *b-3* | 3.89 | 3.40 | 2.89 | 3.27 | 3.82 | 3.33 | 2.83 | 3.17 |
| coronene | *b-1* (bis) | 10.51 | 9.62 | 8.68 | 8.90 | 10.31 | 9.48 | 8.53 | 8.75 |
| | *b-1* (tris) [f] | 14.98 | 14.63 | - | - | 14.71 | 14.35 | - | - |

[a] See Scheme 13 for explicit structures; *b-3* structures are not bicyclobutane forms, but butadienyl derivatives possibly catching *b-2* minima.

[b] For dication, a structure of this type relaxes to the twisted form; for trication, it relaxes to the normal form.

[c] For neutral form, only a flat plateau-point on the potential surface, falling into the *b-3* form.

[d] For trication, only a flat plateau-point on the potential surface, falling into the *b-2* form

[e] For monocation and dication, only flat plateau-points on the potential surface, falling into *b-3* forms; for trication, a structure of this type relaxes to the normal form.

[f] For dication and trication, a structure of this type relaxes to the mono-bicyclobutane form *b-1*.

Table 7. Asymptotic dissociation energies (eV). [a]

| eliminated fragment | PAH | raw | | | | ZPE | | | |
|---|---|---|---|---|---|---|---|---|---|
| | | neutral | monocation | dication | trication | neutral | monocation | dication | trication |
| H$_2$ | coronene | 4.23 | 4.40 | 4.60 | 4.87 | 3.83 | 4.01 | 4.19 | 4.48 |
| | pyrene (1,2) | 4.18 | 4.48 | 4.84 | 5.16 | 3.78 | 4.07 | 4.43 | 4.74 |
| | pyrene (4,5) | 4.33 | 4.54 | 4.81 | 5.17 | 3.92 | 4.13 | 4.39 | 4.75 |
| | pyrene (4,6) | 5.12 | 4.73 | 4.24 | 4.54 | 4.68 | 4.31 | 3.85 | 4.14 |
| H | coronene | 5.10 | 5.59 | 5.09 | 5.55 | 4.74 | 5.26 | 4.73 | 5.20 |
| | pyrene (1) | 5.09 | 5.68 | 5.31 | 4.97 | 4.74 | 5.29 | 4.95 | 4.62 |
| | pyrene (4) | 5.11 | 5.74 | 5.27 | 4.96 | 4.76 | 5.34 | 4.92 | 4.60 |
| | pyrene (5) | 5.09 | 5.49 | 5.36 | 4.92 | 4.73 | 5.11 | 4.99 | 4.55 |
| H + H | coronene | 9.00 | 9.17 | 9.37 | 9.63 | 8.33 | 8.51 | 8.69 | 8.97 |
| | pyrene (1,2) | 8.95 | 9.25 | 9.61 | 9.93 | 8.27 | 8.57 | 8.92 | 9.24 |
| | pyrene (4,5) | 9.10 | 9.31 | 9.57 | 9.94 | 8.42 | 8.62 | 8.88 | 9.25 |
| | pyrene (4,6) | 9.89 | 9.50 | 9.01 | 9.31 | 9.18 | 8.80 | 8.34 | 8.63 |
| C$_2$H$_2$ | coronene | 7.91 | 7.43 | 7.09 | 8.10 | 7.58 | 7.18 | 6.85 | 7.88 |
| | pyrene | 6.80 | 7.08 | 7.56 | 8.05 | 6.49 | 6.76 | 7.23 | 7.74 |
| | pyrene [b] | 7.20 | 6.38 | 6.01 | 7.10 | 6.86 | 6.00 | 5.69 | 6.76 |
| C$_2$H$_4$ | coronene | 8.65 | 8.75 | 9.02 | 9.96 | 8.33 | 8.47 | 8.73 | 9.70 |
| | pyrene | 8.34 | 9.21 | 10.29 | 11.03 | 8.02 | 8.86 | 9.94 | 10.66 |
| C$_2$H$_2^+$ | coronene | | | 7.78 | 3.80 | | | 7.54 | 3.67 |
| | pyrene | | | 6.70 | 2.08 | | | 6.41 | 1.87 |
| | pyrene [b] | | | 6.00 | 0.53 | | | 5.65 | 0.34 |
| C$_2$H$_4^+$ | coronene | | | 8.17 | 4.80 | | | 7.82 | 4.54 |
| | pyrene | | | 7.90 | 3.89 | | | 7.50 | 3.57 |

[a] See Scheme 16 for labeling.   [b] This line refers to fragmentation of type 2 (see Scheme 20).

Table 8. Barriers for concerted H$_2$ eliminations *via* 1,2-hydrogen shifted minima (eV).

| system | charge | from H-shifted | | overall | |
|---|---|---|---|---|---|
| | | raw | ZPE | raw | ZPE |
| coronene | 0 | 1.83 | 1.64 | 5.35 | 5.06 |
| | 1 + | 2.97 | 2.69 | 5.32 | 5.04 |
| | 2 + | 2.52 | 2.30 | 5.27 | 4.99 |
| | 3 + | 2.88 | 2.27 | 5.17 | 4.87 |
| pyrene | 0 | 1.89 | 1.69 | 5.29 | 5.00 |
| | 1 + | 2.80 | 2.54 | 5.42 | 5.10 |
| | 2 + | 2.14 | 1.91 | 5.38 | 5.03 |
| | 2 + [a] | 2.89 | 2.65 | 5.13 | 4.81 |
| | 3 + | 2.92 | 2.59 | 5.09 | 4.72 |

[a] Pathway *via* the 5,6-hydrogen shifted minimum, for comparison.

Table 9. Calculated barriers for direct elimination of $C_2H_2$ from normal forms. [a]

| system | charge | raw | | ZPE | |
|---|---|---|---|---|---|
| | | TS | DE | TS | DE |
| coronene | 0 | 9.12 | 7.91 | 8.86 | 7.58 |
| | 1 + | 8.05 | 7.43 | 7.86 | 7.18 |
| | 2 + | 7.22 | 7.09 | 7.05 | 6.85 |
| pyrene | 0 | | 6.80 | | 6.49 |
| | 1 + | 7.61 | 7.08 | 7.35 | 6.76 |
| | 2 + | 7.15 | 6.70 | 6.93 | 6.41 |

[a] In eV; TS: transition state energy; DE: dissociation energy; for pyrene dication, the eliminated fragment is $C_2H_2^+$.

Table 10. Calculated barriers for direct concerted elimination of $C_2H_4$ from 'prepared' double hydrogen-shifted isomers. [a]

|  |  | raw | | | ZPE | | |
|---|---|---|---|---|---|---|---|
| system | charge | -$CH_2$-$CH_2$- | TS | DE | -$CH_2$-$CH_2$- | TS | DE |
| coronene | 0 | 5.00 | 9.18 | 8.65 | 4.92 | 8.91 | 8.33 |
|  | 1 + | 4.76 | 9.01 | 8.75 | 4.72 | 8.77 | 8.47 |
|  | 2 + | 5.31 | 8.78 | 8.17 | 5.21 | 8.51 | 7.82 |
|  | 3 + | 4.95 |  | 4.80 | 4.89 |  | 4.54 |
| pyrene | 0 | 6.62 | 8.87 | 8.34 | 6.48 | 8.61 | 8.02 |
|  | 1 + | 5.21 | 9.12 | 9.21 | 5.10 | 8.81 | 8.86 |
|  | 2 + | 5.12 | 9.56 | 7.90 | 4.94 | 9.23 | 7.50 |
|  | 3 + | 4.45 | 8.40 | 3.89 | 4.31 | 8.08 | 3.57 |

[a] in eV; -$CH_2$-$CH_2$-: double hydrogen-shifted minimum; TS: transition state energy; DE: dissociation energy; for dications and trications, the eliminated fragment is $C_2H_4^+$.

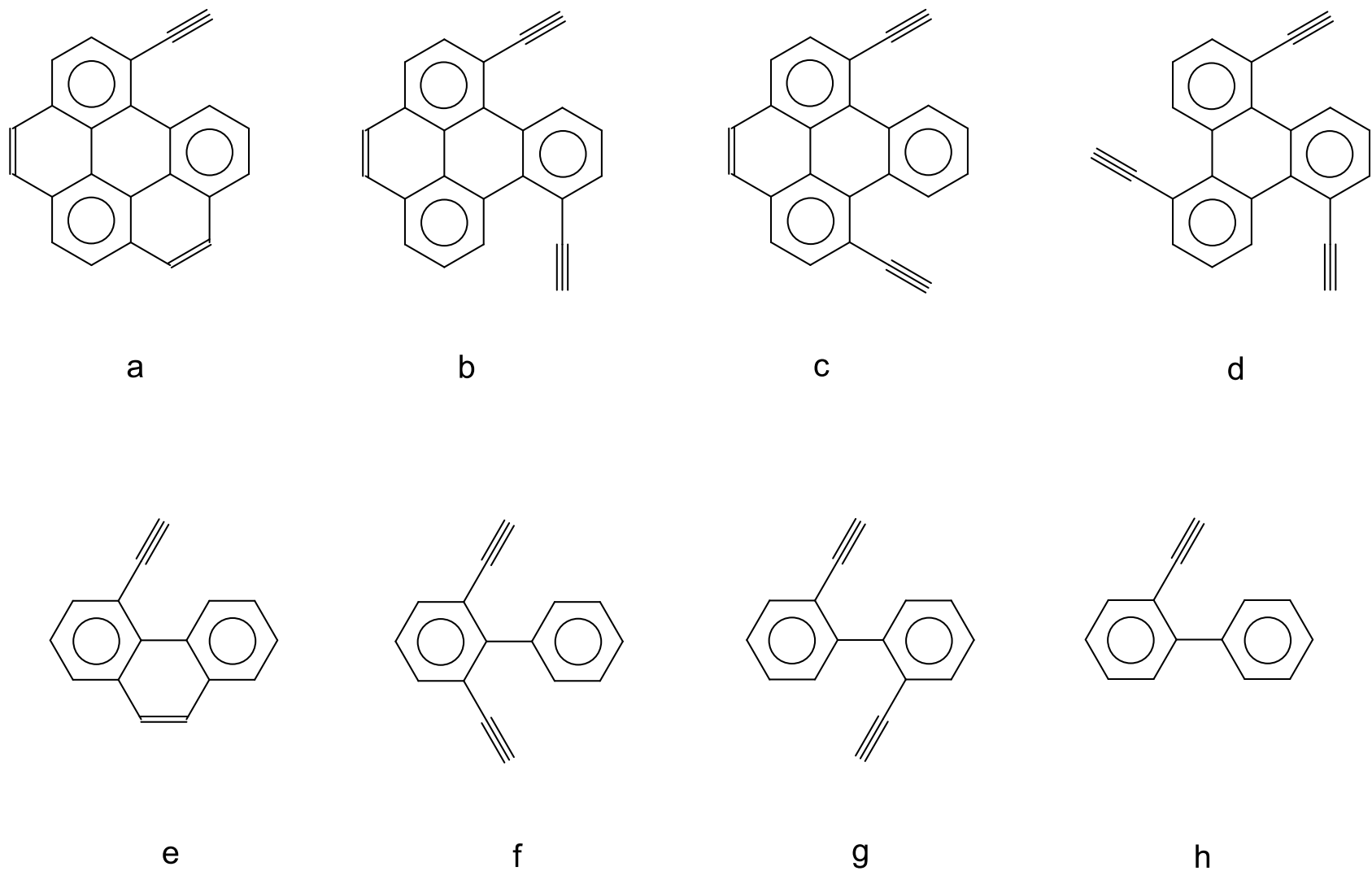

Figure 1. Various ways of opening rings into extracyclic ethynyl derivatives in coronene (a-d), pyrene (e-g), and phenanthrene (h).

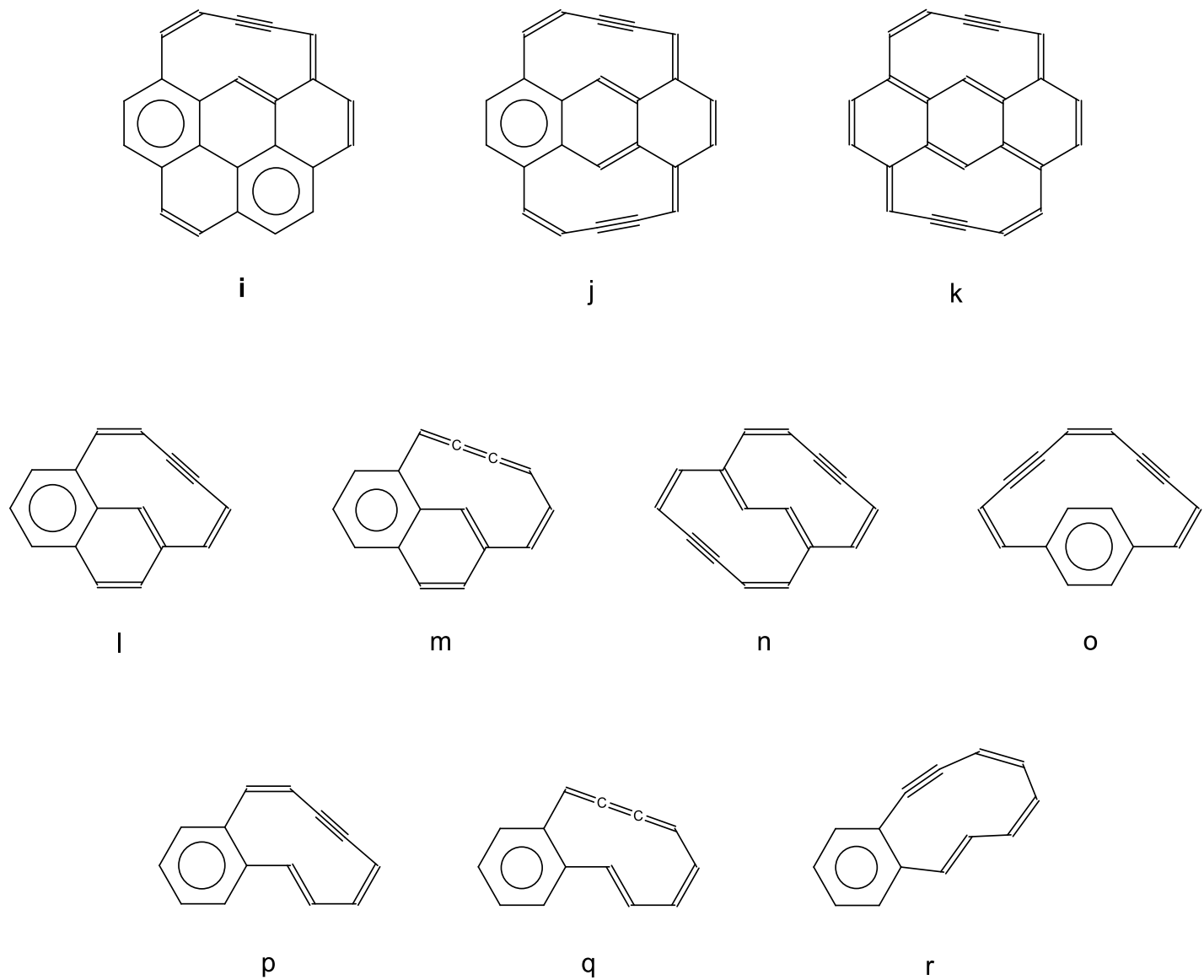

Figure 2. Some cases of ring opening into 2-butyne derivatives of coronene (i-k), pyrene (l-o), and phenanthrene (p-r).

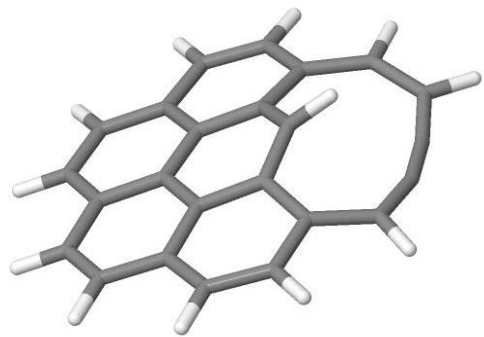 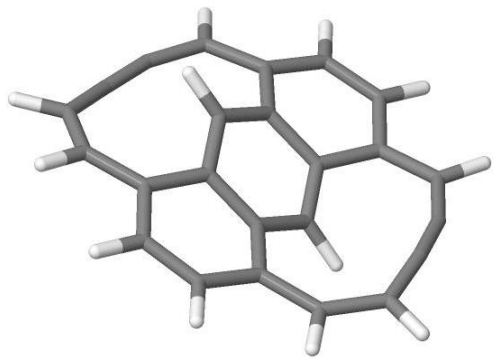 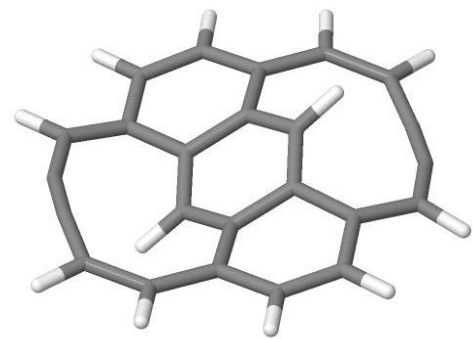

i  j  k

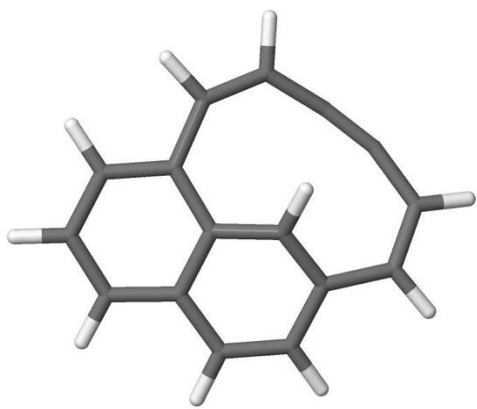 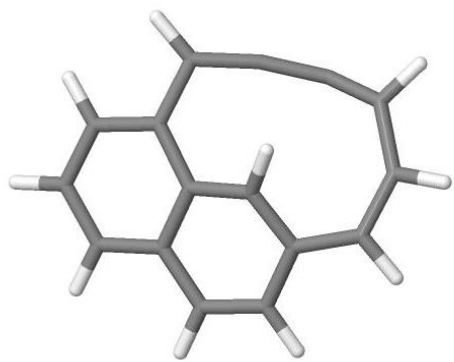 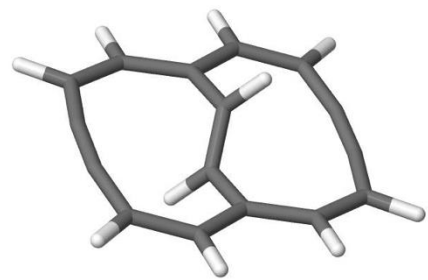 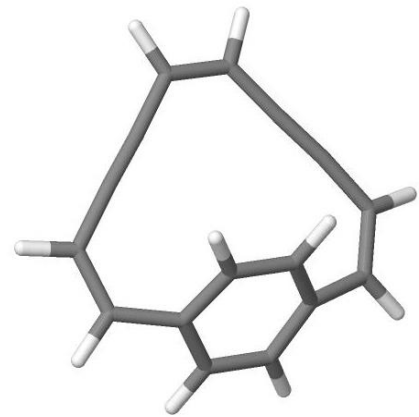

l  m  n  o

Figure 3. Optimized 2-butynyl derivatives of coronene and pyrene.

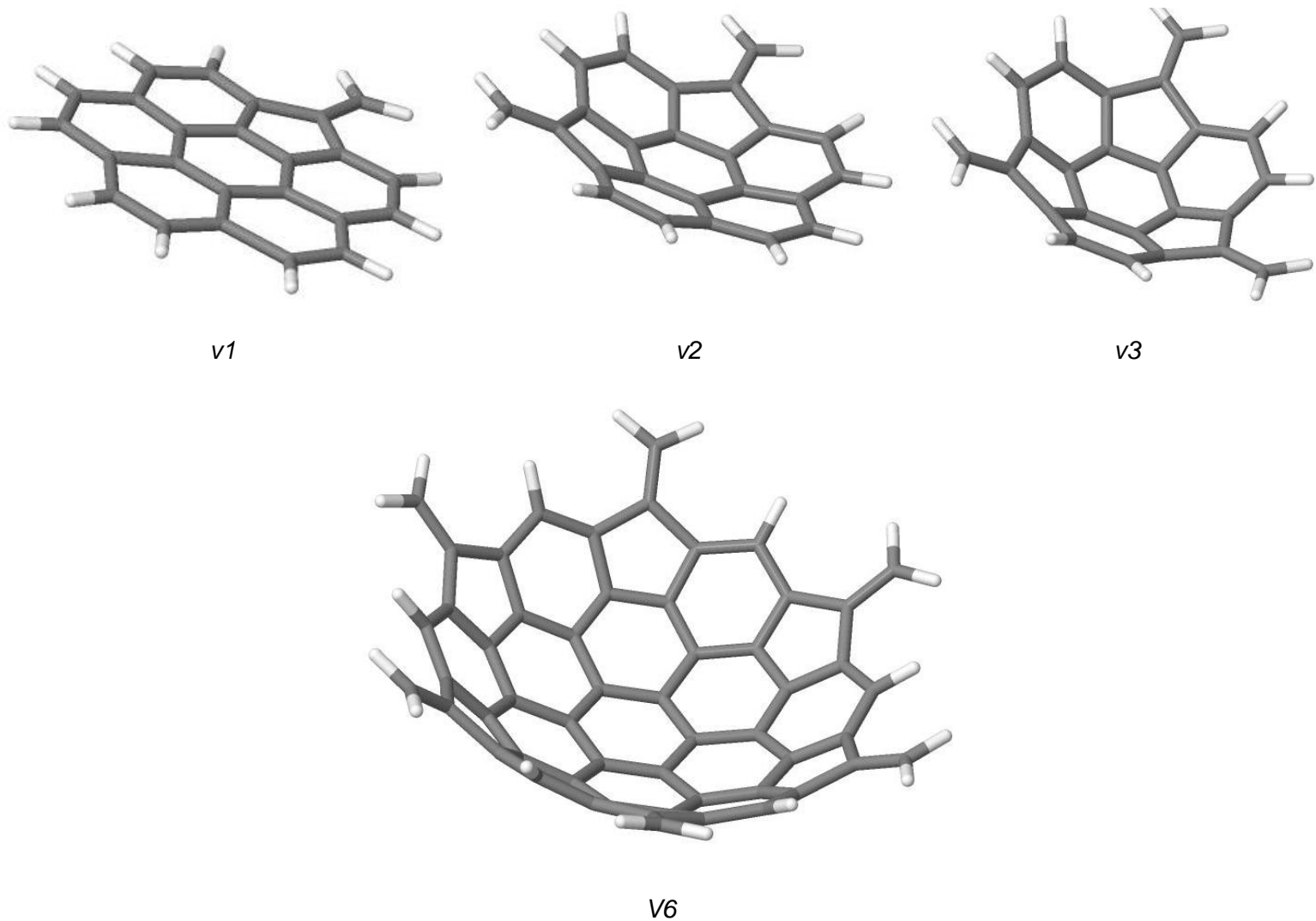

*v1*  *v2*  *v3*

*V6*

Figure 4. Optimized geometries of vinylidene forms for coronene (top) and circum-coronene (bottom).

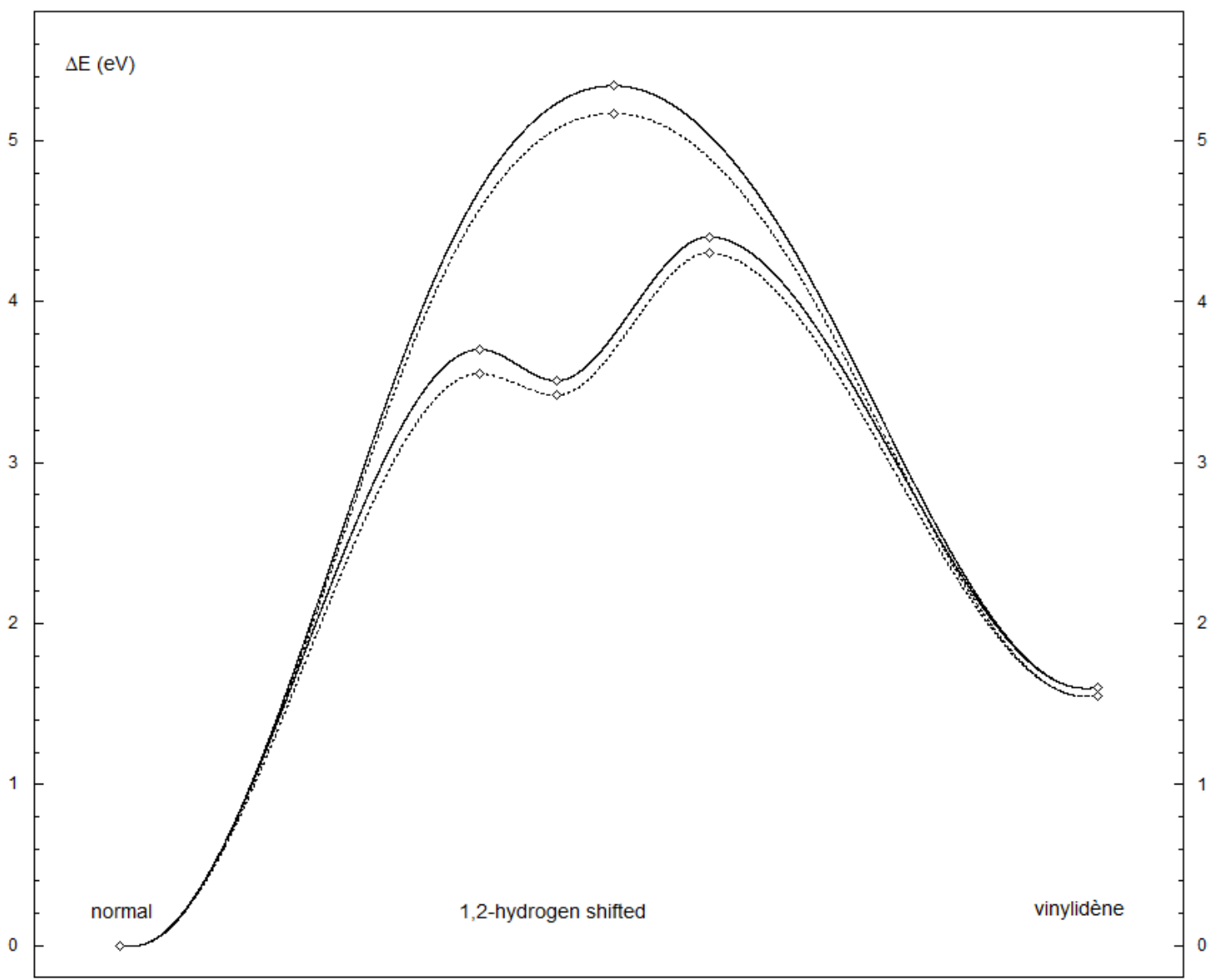

Figure 5. Schematic energy profiles for the two pathways linking the normal form of neutral coronene to its vinylidene bridged form. Full curves: raw calculation; dashed curves: after ZPE correction

Figure 6. Multiple views of CAS-calculated structures of twisted forms for neutral coronene

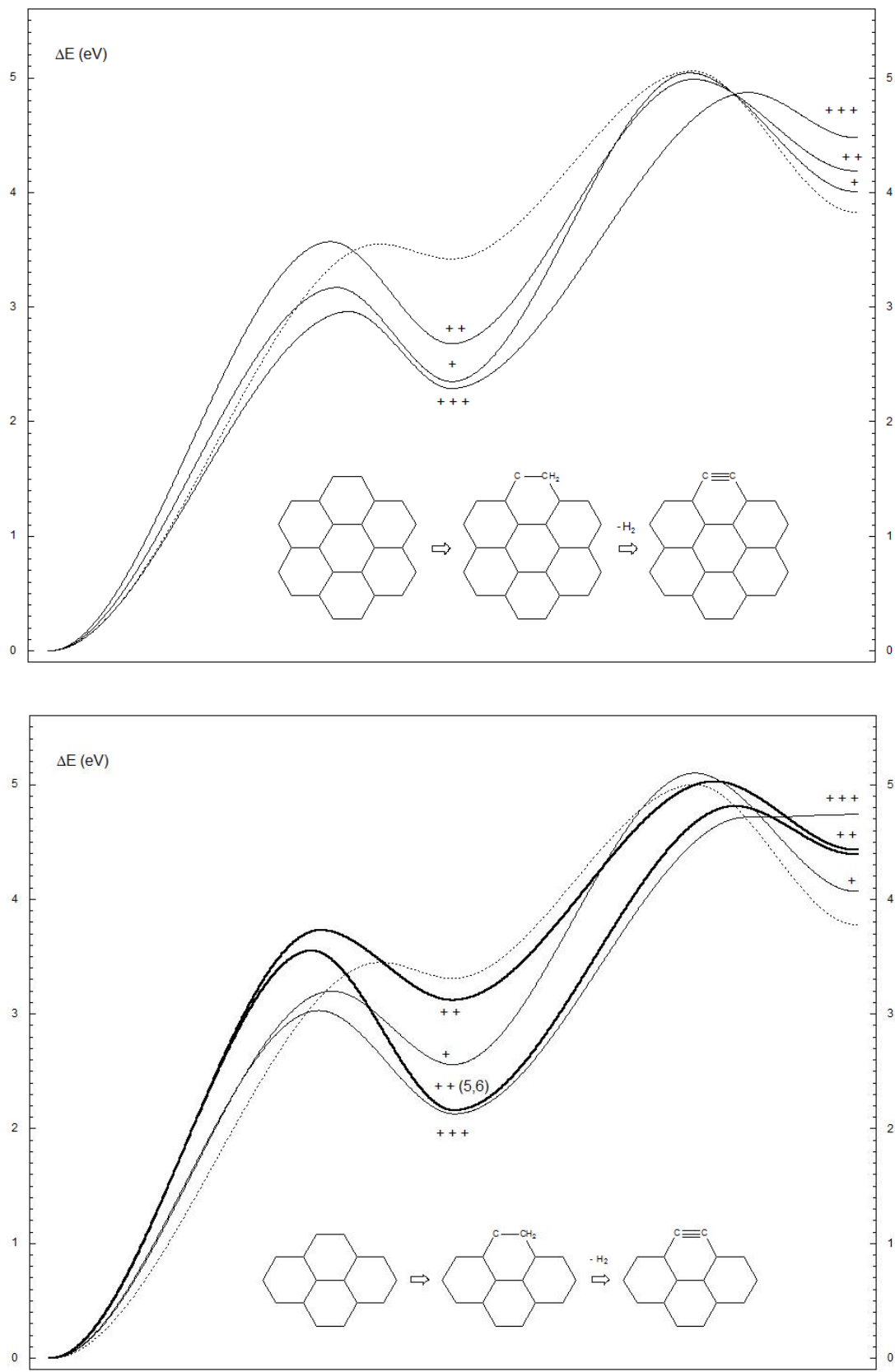

Figure 7. Schematic energy profile (ZPE level) for H$_2$ elimination from coronene (top) and pyrene (bottom) *via* 1,2-hydrogen shifted isomers. Dashed curves corresponds to neutral states. Elimination from 5,6 positions of pyrene dication is given for comparison.

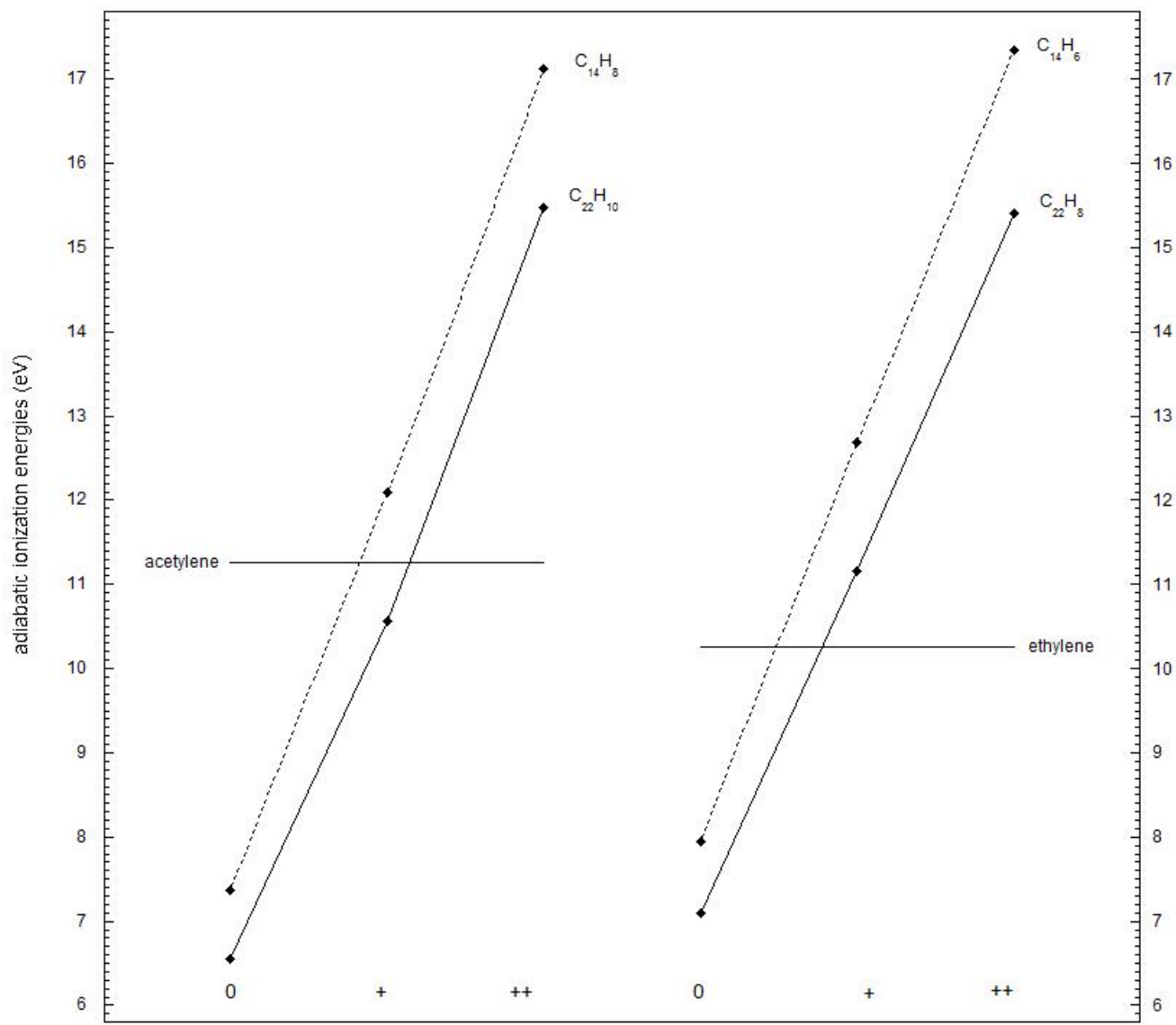

Figure 8. Adiabatic ionization energies of remaining fragments after elimination of acetylene (left) or ethylene (right) from coronene (full lines) and pyrene (dashed lines) in neutral and charged forms. The horizontal lines mark adiabatic ionization energies of acetylene and ethylene (all values at ZPE level).

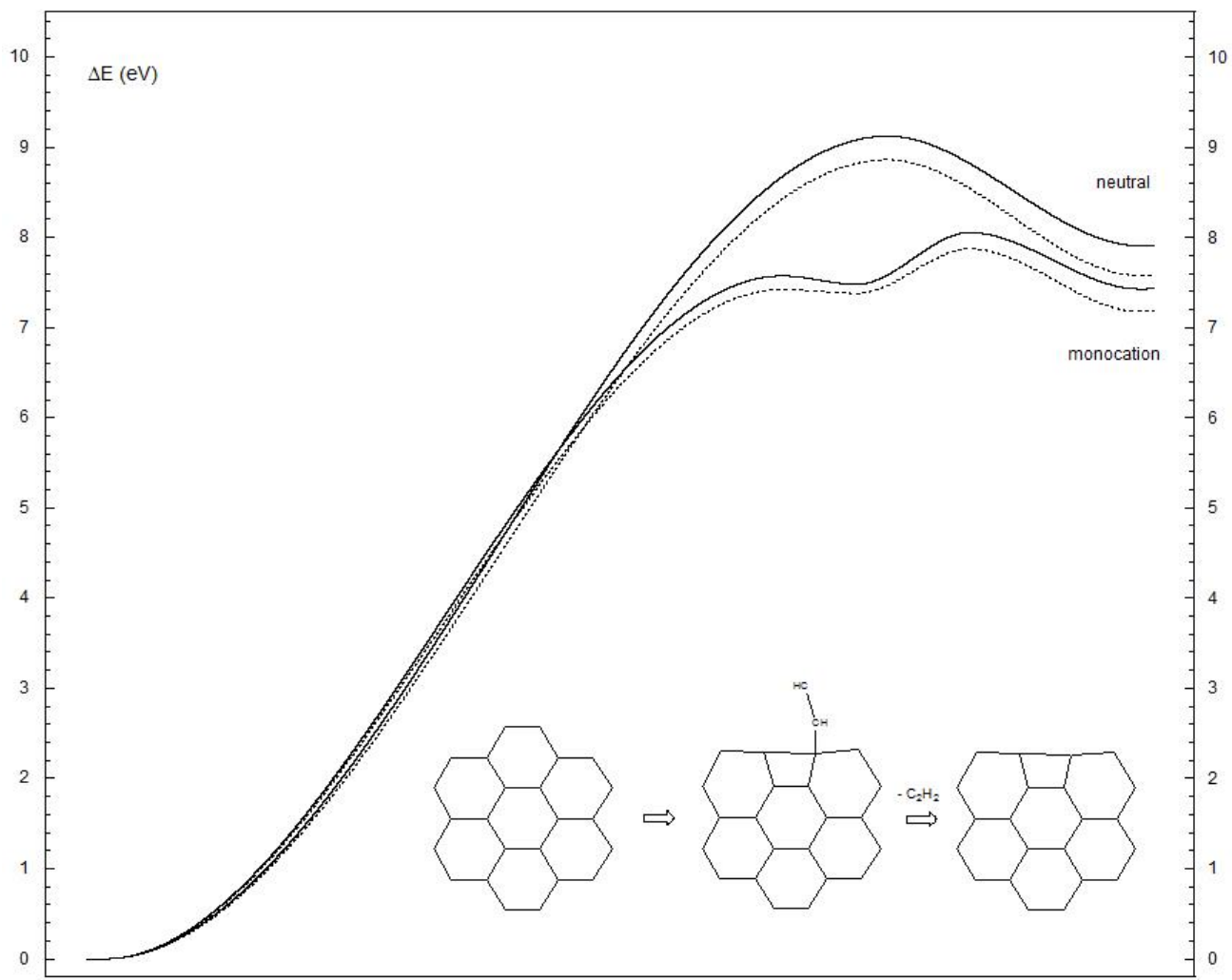

Figure 9. Schematic energy profile for acetylene elimination from neutral and cationic coronene (full curve: raw; dashed curve ZPE).

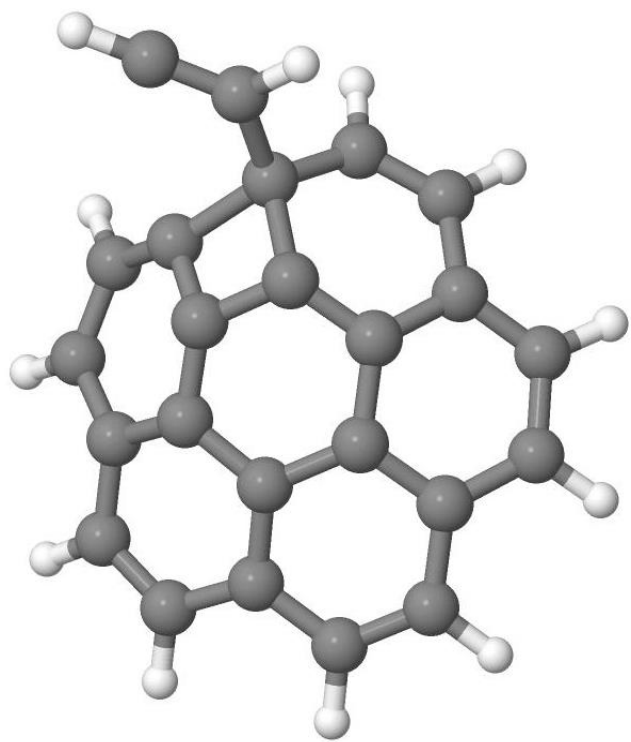 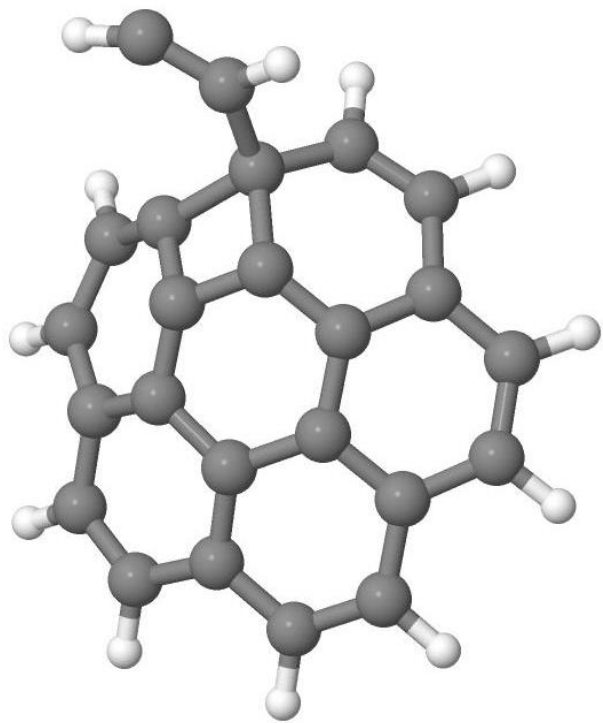 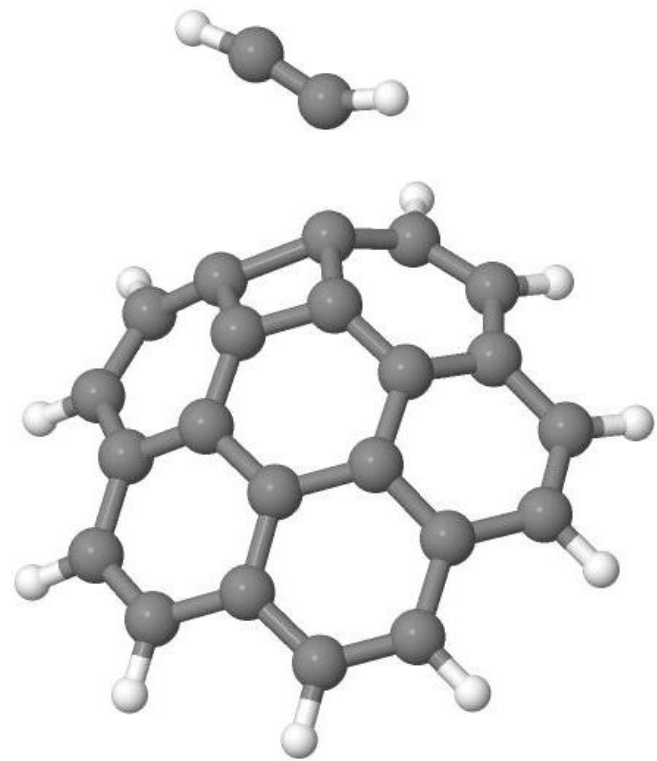

TS' (562 $i$ cm$^{-1}$)            local minimum            TS (612 $i$ cm$^{-1}$)

Figure 10. Stationary points along the acetylene dissociation path in coronene monocation.

**Appendix A.** Allowed (top, middle) and forbidden (bottom) di-allenic ring opening in PAHs.

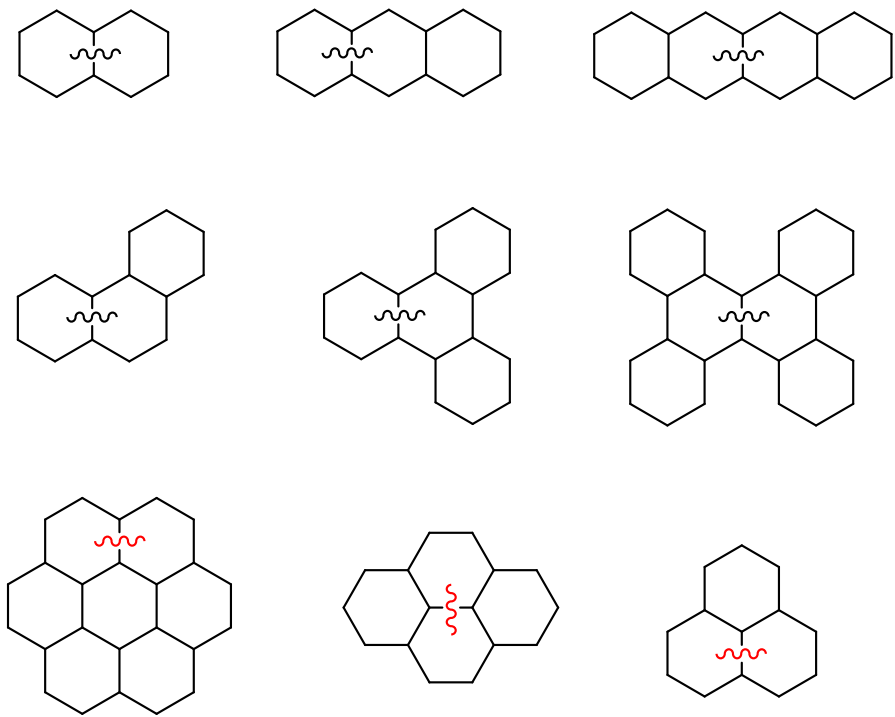

**Appendix B**. Relative energies of di-allenic opened forms with respect to normal forms.

|  | ΔE (eV) | |
|---|---|---|
|  | raw | ZPE |
| naphthalene | 4.54 | 4.38 |
| naphthalene (TS) | 5.52 | 5.23 |
| phenanthrene | 5.28 | 5.12 |
| tetra-benzo-naphthalene | 6.98 | 6.78 |

**Appendix C**. Energy increments required for successive ring openings into extracyclic-ethynyl forms for neutral species (see Figure 1 for labeling).

|  |  | ΔE (eV) | |
|---|---|---|---|
|  |  | raw | ZPE |
| coronene | mono-ethynyl | 2.71 | 2.60 |
|  | di-ethynyl (b) | 2.62 | 2.52 |
|  | di-ethynyl (c) | 2.64 | 2.54 |
|  | tri-ethynyl (d) | 2.56 | 2.46 |
| pyrene | mono-ethynyl | 2.47 | 2.36 |
|  | di-ethynyl (*g*) | 2.17 | 2.06 |
|  | di-ethynyl (f) | 2.13 | 2.02 |
| phenanthrene | mono-ethynyl | 2.28 | 2.17 |

**Appendix D.** Energy barriers (eV) along the two paths to poly-vinylidene isomers.

| | | raw | | ZPE | |
|---|---|---|---|---|---|
| system | destination | overall | relative | overall | relative |
| neutral coronene (direct) | mono-vinylidene | 5.34 | 5.34 | 5.17 | 5.17 |
| | di-vinylidene | 7.32 | 5.72 | 7.09 | 5.54 |
| | tri-vinylidene | 9.15 | 5.57 | 8.88 | 5.40 |
| neutral pyrene (direct) | mono-vinylidene | 4.91 | 4.91 | 4.75 | 4.75 |
| | di-vinylidene | 7.04 | 5.88 | 6.81 | 5.70 |
| cationic pyrene (direct) | mono-vinylidene | 5.20 | 5.20 | 4.96 | 4.96 |
| | di-vinylidene | 7.26 | 5.81 | 6.98 | 5.60 |
| neutral pyrene (indirect) | H-shifted (TS1) | 3.60 | 3.60 | 3.45 | 3.45 |
| | mono-vinylidene (TS2) | 4.22 | 0.82 | 4.12 | 0.81 |
| | H-shifted mono-vinylidene (TS1) | 4.75 | 3.59 | 4.55 | 3.44 |
| | di-vinylidene (TS2) | 6.01 | 1.51 | 5.84 | 1.48 |
| cationic pyrene (indirect) | H-shifted (TS1) | 3.36 | 3.36 | 3.20 | 3.20 |
| | mono-vinylidene (TS2) | 3.79 | 1.17 | 3.69 | 1.13 |
| | H-shifted mono-vinylidene (TS1) | 4.47 | 3.02 | 4.27 | 2.89 |
| | di-vinylidene (TS2) | 5.72 | 2.04 | 5.55 | 1.97 |

**Appendix E**. Origin of twisted forms and computational requirements.

To understand the origin of the twisted forms, one can start from the *cis-trans* isomerization of secondary olefins, the simplest representative of which is 2-butene H$_3$C–HC=CH–CH$_3$. Exploring such a torsional pathway along the CCCC dihedral-angle coordinate (ω), for instance at simple CAS(2,2) level, there exists two separate valleys associated with the *cis* and *trans* starting configurations. Stepping onto one of these paths, each olefinic carbon pyramidalizes while keeping the H–C=C–H dihedral angle in its starting configuration, i.e. 0° from *cis* configuration, 180° from *trans* configuration. This holds until a torsional angle threshold is reached, around ω≈70° when starting from *trans*, and ω≈110° when starting from *cis*, beyond which the double-bond configuration switches and the energy collapse onto the other form. Suppose you twist the *trans* form up to an ω value around 75°. At this point, lying about 2 eV above the normal relaxed form, the HCCH arrangement is still *trans* (see below), and introducing appropriate bridging between the ends of butene skeleton (red clip) would prevent the system to return both to *trans* arrangement due to the cycle (clip), and to

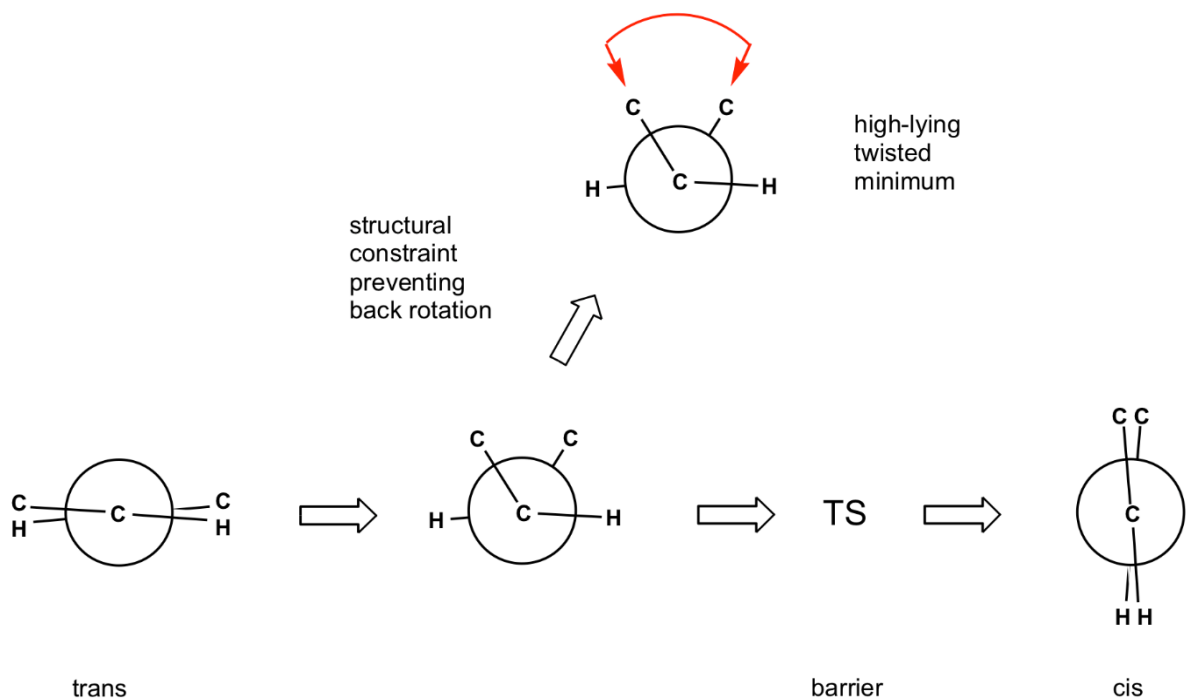

*cis* arrangement as long as we keep before the barrier. In this way, the system is caught in a high-lying local minimum, characterized by *trans* arrangement at H–C=C–H, with a C-C-C-C dihedral angle ω of 70-80°. In a way, this twisted form can be considered as a virtual minimum in butene, while it becomes a real minimum in a PAH, the whole skeleton acting here as the blocking clip. In this twisted form, the double-bond π electrons, formerly in coplanar 2p$_z$ carbon orbitals, are now in two sp$^3$-type hybrids which are skewed by about 60°, thus preserving residual overlap and possible conjugation with the remaining polycyclic environment when in PAH context. In like manner, most cycloalkenes are expected to possess such high-lying local minimum on the ground-state closed-shell singlet surface, lying at 2-4 eV above the regular form, depending on environment-driven deconjugation and geometric strains.

The question of accurate description for such systems bearing a twisted deconjugated π bond immediately arises, both for assessing their energy heights, and for estimating the barrier to overcome for back collapsing to their normal conjugated forms. Indeed, from Hammond's postulate, one should expect for these high-lying minima weak backwards barriers, with corresponding structures close to these twisted forms. In other words, do such twisted forms really exist, and to which extent are they viable species ? In the present PAH explorations, geometries of this type are found rather easily at DFT level, lying around 3-4 eV above planar ground states. However, the B3LYP hybrid functional may overestimate delocalization effects between carbon *sp$^3$* hybrid orbitals twisted at about 60°. Such configurations have some diradical character and should be better described through MCSCF CAS(2,2) procedures. However, the planar ground-state reference may require more than CAS(2,2), as sets of degenerate orbitals may be involved in ground states of these polycyclic systems. Typically, CAS(2,2) should be sufficient to estimate the energies of the two forms of cyclohexene, while CAS(4,4) or CAS (6,6) should be required to estimate them in benzene.

**Appendix F**. Schematic energy profiles for pathways linking normal dicationic coronene to its twisted form, and to its bicyclobutane form *b-1* (top), and normal tricationic coronene to its bicyclobutane form *b-1* (bottom). Full curves: raw calculation; dashed curves: after ZPE corrections.

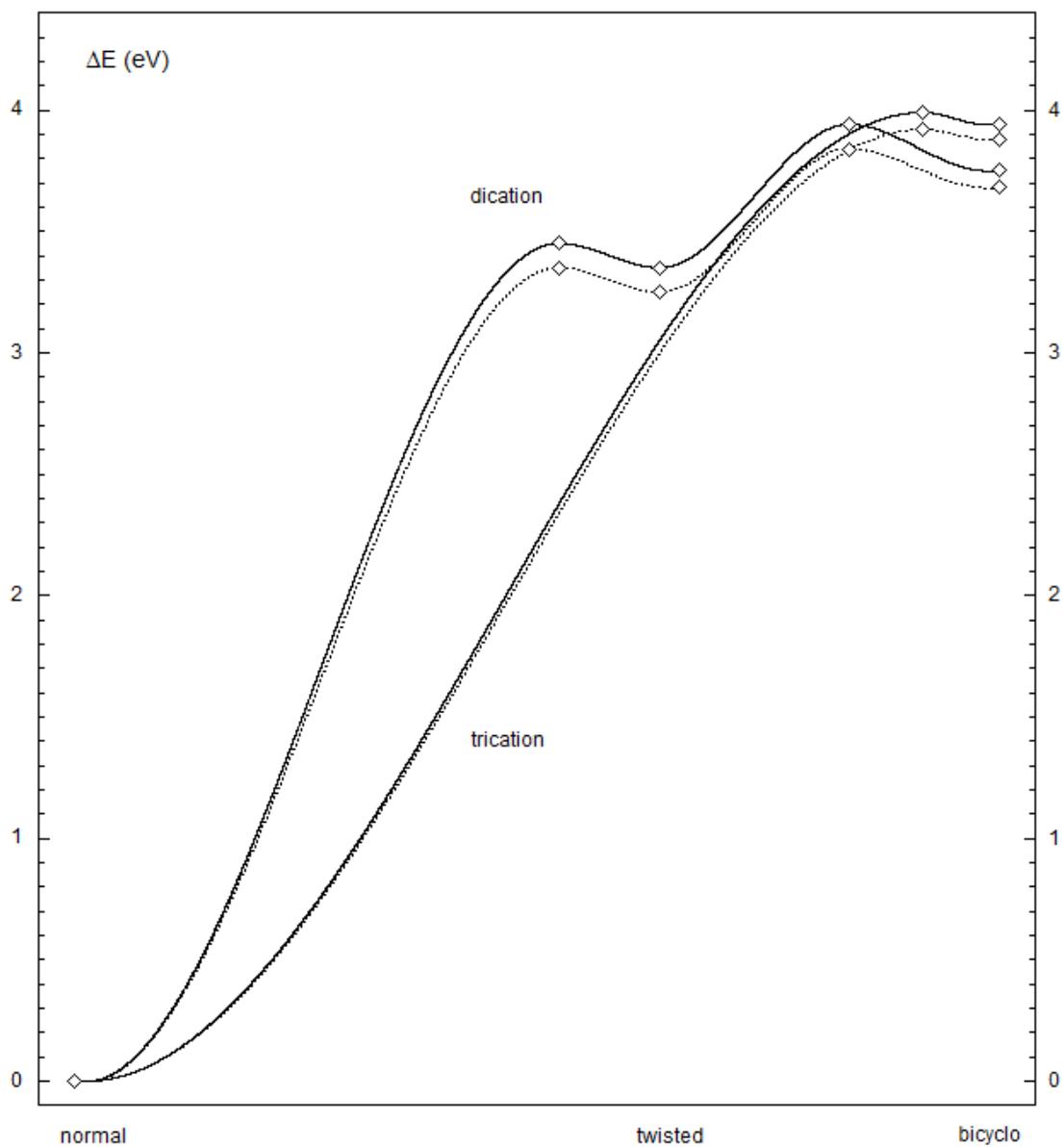

**Appendix G**. Definition of bicyclobutane-type structures (top row), with corresponding conjugated frames.

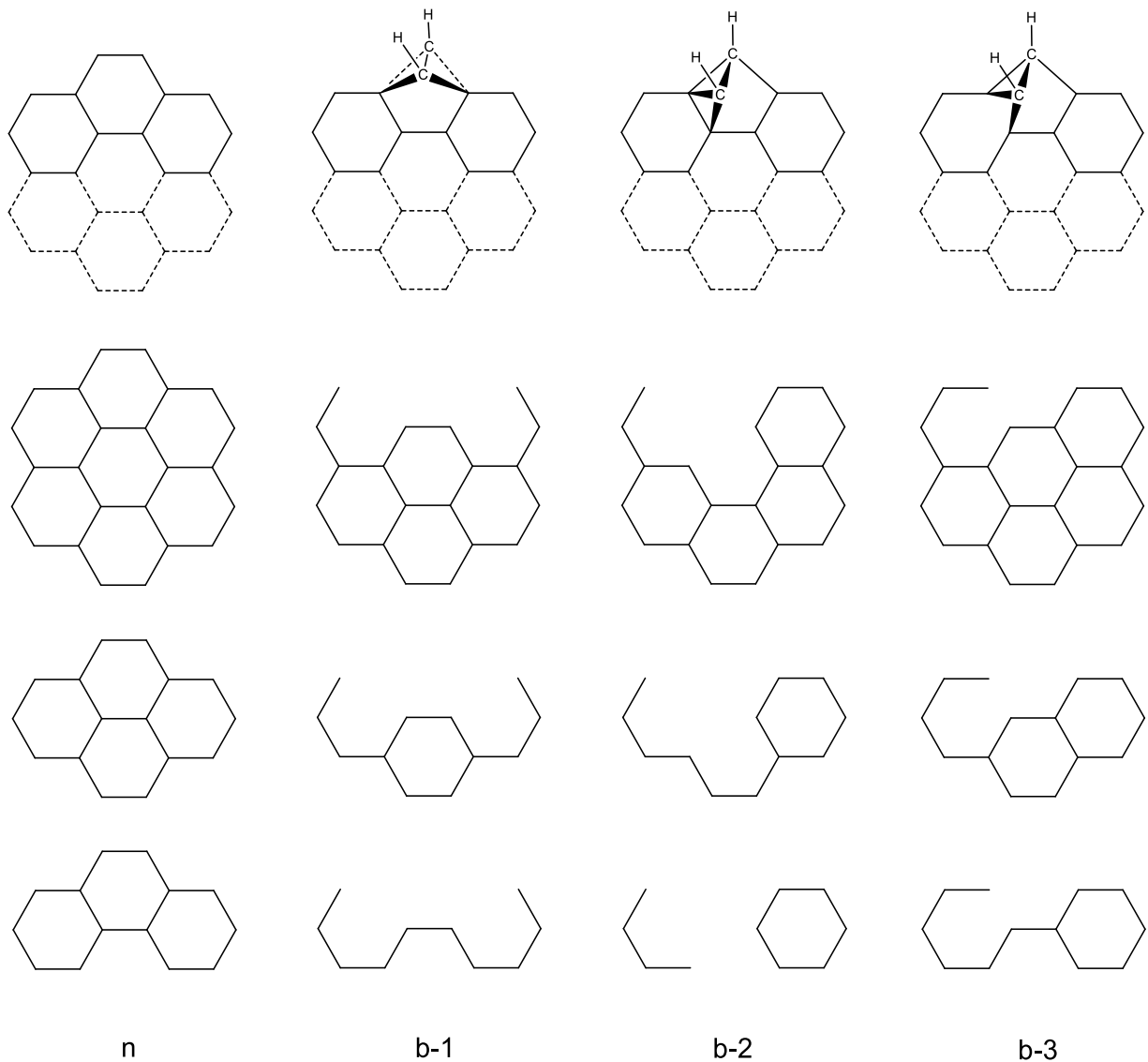

| n | b-1 | b-2 | b-3 |

**Appendix H**. Calculated relative energies along selected rearrangement pathways. [a]

| PAH | charge | connection | raw | ZPE |
|---|---|---|---|---|
| coronene | neutral | twisted | 3.94 | 3.87 |
| | | TS | 5.75 | 5.59 |
| | | *b-2* | 4.91 | 4.83 |
| | cation | normal | 0. | 0. |
| | | TS | 4.68 | 4.61 |
| | | *b-2* | 4.58 | 4.54 |
| | di-cation | twisted | 3.48 | 3.46 |
| | | TS | 3.94 | 3.84 |
| | | *b-1* | 3.75 | 3.68 |
| | | twisted | 3.48 | 3.46 |
| | | TS | 4.39 | 4.32 |
| | | *b-3* | 3.94 | 3.93 |
| | tri-cation | normal | 0. | 0. |
| | | TS | 3.99 | 3.92 |
| | | *b-1* | 3.94 | 3.88 |
| | | normal | 0. | 0. |
| | | TS | 4.37 | 4.33 |
| | | *b-3* | 4.27 | 4.27 |
| pyrene | neutral | twisted | 3.68 | 3.61 |
| | | TS | 5.61 | 5.46 |
| | | *b-3* | 4.35 | 4.28 |
| | cation | normal | 0. | 0. |
| | | TS | 4.71 | 4.57 |
| | | *b-2* | 4.41 | 4.33 |
| | di-cation | *b-1* | 3.75 | 3.64 |
| | | TS | 4.34 | 4.19 |
| | | *b-2* | 4.05 | 3.97 |

[a] In eV; see Scheme 13 for labeling.

**Appendix I**. Possible scrambling *via* vinylidene minima; left: direct mechanism; right: indirect mechanism.

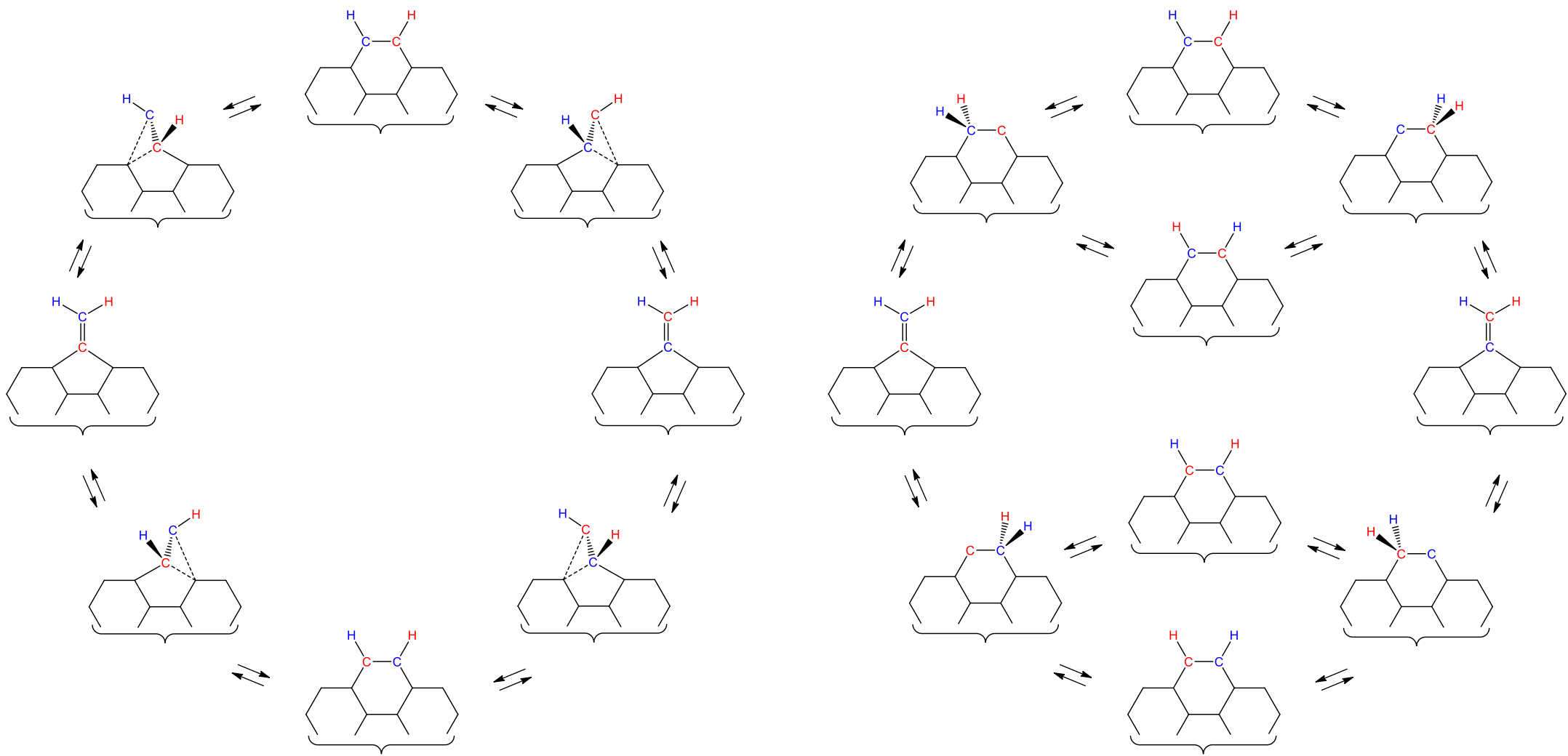